\documentclass[twocolumn,preprintnumbers,pre]{revtex4}
\usepackage{graphicx, subfigure}
\usepackage{amssymb, amsmath,amssymb,amsfonts}
\usepackage{amsthm,mathrsfs,amsopn}
\usepackage{dcolumn}
\usepackage{bm}
\usepackage{ulem}
\usepackage{enumerate}
\usepackage{color}
\usepackage[utf8]{inputenc}
\usepackage{mathtools}
\usepackage{tikz}
\usepackage{comment}
\usepackage{hyperref}
\usepackage{orcidlink}
\graphicspath{{./Figures}}
\theoremstyle{plain} 

\begin{document}

\title{GDP-Driven Structural and Dynamical Heterogeneity in the Synchronization of Chaotic Macroeconomic Networks}

\author{Thierry Njougouo\,\orcidlink{0000-0001-7706-7674}}
\email{thierry.njougouo@imtlucca.it}
\affiliation{$^1$IMT School for Advanced Studies, Piazza San Francesco 19, 55100 Lucca, Italy}

\author{Fernando Fagundes Ferreira\,\orcidlink{0000-0001-9183-8287}}
\affiliation{$^2$FFCLRP - Faculdade de Filosofia, Ciências e Letras - Universidade de Sao Paulo - USP-RP, Av. Bandeirantes, 3900 - Bairro Monte 
Alegre, Ribeirão Preto 14040-901, São Paulo, Brazil}

\author{Diego Garlaschelli\,\orcidlink{0000-0001-6035-1783}}
\affiliation{$^1$IMT School for Advanced Studies, Piazza San Francesco 19, 55100 Lucca, Italy}
\affiliation{$^3$Instituut-Lorentz for Theoretical Physics, Leiden University, 2333 CA Leiden, The Netherlands}

\begin{abstract}
We investigate the emergence of synchronization in a network of coupled chaotic macroeconomic systems. Each node represents an economy characterized by three key variables savings, gross domestic product (GDP), and foreign capital inflows. These economies interact or are connected through a fitness-based probability that depends on the potential GDP of each node. This formulation allows both structural heterogeneity, arising from uneven network connectivity, and dynamical heterogeneity, due to differences in local parameters, to be explored within a unified framework. Using both numerical simulations and a mean-field approximation, by varying the coupling strength and the degree of heterogeneity of both network topology and dynamical behavior of the nodes, we analyze synchronization transitions.  Our results show that the mean-field approach accurately captures the collective dynamics in homogeneous and fully connected networks even with heterogeneity within the intrinsic dynamic of the nodes but fails when strong heterogeneity in the structure of the network is introduced. In heterogeneous networks, the system exhibits partial synchronization and on--off intermittency, where coherent phases of global synchronization alternate with abrupt desynchronization bursts. The distribution of laminar phase durations follows a power-law scaling, consistent with theoretical predictions for intermittent synchronization. From an economic perspective, these results suggest that global business cycle synchronization is inherently fragile: strong integration can promote temporary coordination among economies, but structural and dynamical disparities inevitably lead to intermittent breakdowns of collective behavior.
\end{abstract}

\maketitle

\section{Introduction}\label{sec::int}

Complex networks provide a unifying framework for describing systems composed of many interacting entities connected through homogeneous or heterogeneous patterns of interaction. Unlike the simple and artificial case of regular graphs, complex networks often exhibit nontrivial structural features such as small-world connectivity~\cite{watts1998collective}, scale-free degree distributions~\cite{barabasi2003scale}, random connectivity~\cite{erdds1959random} and modular or hierarchical organization~\cite{rives2003modular}. Structural properties strongly influence the collective dynamics that emerge on the networks, including consensus~\cite{reina2024speed,njougouo2023heterogeneous,meylahn2024opinion}, chimera state~\cite{bera2017chimera,simo2021chimera,muolo2024phase}, contagion~\cite{perez2022network,cencetti2023distinguishing,saberi2020simple}, critical transitions in synchronization~\cite{arola2022emergence,njougouo2020effects,njougouo2020dynamics,zheng2000generalized}, etc. These findings have motivated the applications of complex network theory to a wide variety of real-world systems, where similar structural–dynamical interplays are observed and used, among them, we can list biology and neuroscience, engineering, sociology, and economics.

Since the advent of globalization, complex networks have emerged as a powerful tool for modeling and analyzing global economic interactions, providing deeper insights into the  structural and dynamical organization of economic systems. From the perspective of complex network theory, economic and financial interactions naturally form networked structures composed of a large number of heterogeneous nodes representing agents, firms, or countries, with links encoding flows of goods, capital, or information~\cite{schweitzer2009economic,battiston2016complexity}. The topology of these networks plays a fundamental role in shaping macroeconomic outcomes, influencing how shocks propagate through the different microeconomies of the underline network, how synchronization among business cycles emerges, and how systemic crises develop through the whole macroeconomic system~\cite{haldane2011systemic,acemoglu2012network}. Further illustration of the impact of topology on economic networks is investigated in Refs.~\cite{fagiolo2010evolution,bardoscia2021physics} showing that global trade and financial networks exhibit small-world and scale-free properties. Such topological features enhance efficiency in the exchange of goods and capital but also amplify contagion during crises~\cite{haldane2011systemic}.

From the complex systems theory perspective, an economy can be analytically modeled as a nonlinear dynamical unit interacting within a network of coupled systems. To formalize this idea, we consider each economy as a nonlinear and chaotic dynamical system whose internal variables such as savings, gross domestic product (GDP), and foreign capital inflows evolve through mutual interactions as introduced by Bouali~\cite{bouali1999feedback}. This framework, allows us to consider the GDP as a double indicator, first as an indicator of internal economic performance and then, as a connection channel through which global interdependencies emerge. By embedding these dynamical units within a network whose topology encodes the economic relationships, we can investigate how structural heterogeneity and dynamical heterogeneity of the nodes jointly shape the onset of collective behaviors. This double heterogeneity is explained by the fact that real economic networks are both structurally heterogeneous---due to uneven trade and financial linkages---and dynamically heterogeneous, as economies differ in productivity, growth potential, and internal stability~\cite{battiston2016complexity,colon2017economic}. The interplay between these two sources of heterogeneity is central to understanding the robustness of the global economic coordination.

In this paper, we investigate the emergence of synchronization in a network of coupled chaotic macroeconomic systems. Each node in the network represents an economy described by a three-dimensional chaotic systems involving savings, GDP, and foreign capital inflows~\cite{camargo2022synchronization}. The network topology is defined by a fitness-based probabilistic connection function that depends on the potential GDP of each economy/node, allowing us to modulate structural heterogeneity and homogeneity through a single fitness parameter~\cite{garlaschelli2004fitness}. By varying both the coupling strength and the degree of heterogeneity of the network, we explore the limits of the mean-field approximation and examine the transition to synchronization in different scenarios.

Our results illustrate that the mean-field approximation provides an accurate description of collective dynamics in homogeneous systems on fully connected networks, but fails in the presence of strong structural heterogeneity. In the case of heterogeneous dynamical systems even in fully connected networks, synchronization emerges intermittently, with coherent laminar phases interrupted by bursts of desynchronization. The durations of these laminar phases follow a power-law distribution, a hallmark of on--off intermittency. From an economic interpretation, these results suggest that global coordination among economies is intrinsically fragile: even under strong coupling, synchronized growth phases are transient and punctuated by abrupt divergences triggered by internal or external perturbations.

The paper is organized as follows. Sec.~\ref{sec::sec2} presents the dynamics of a single macroeconomic system in the absence of any interaction. Sec.~\ref{sec::sec3} introduces the construction of the economic network based on the potential GDP and analyzes the synchronization transitions. The analysis is carried out first for homogeneous dynamical systems on both homogeneous and heterogeneous network topologies, and then for heterogeneous dynamical systems under the same structural configurations. Finally, Sec.~\ref{sec::con} discusses the implications of these findings for global economic coordination and concludes the paper.

\section{Model Description of a Single Macroeconomical System }\label{sec::sec2}

Let us consider the macroeconomic system defined by Eq.~\ref{eq::Cam}, which is an extension of the Bouali~\cite{bouali1999feedback} model proposed by Camargo et al \cite{camargo2022synchronization}. 
\begin{equation}
    \begin{cases}
        \dot{x} = my + px(d - y^2), \\
        \dot{y} = vy + wx + cz, \\
        \dot{z} = sx - ry,
    \end{cases}
    \label{eq::Cam}
\end{equation}
This macroeconomic model considered here describes the coupled dynamics of three interacting variables: savings $x$, gross domestic product (GDP) $y$, and foreign capital inflow $z$. The parameters governing their mutual feedbacks, which determine the nonlinear evolution of the system, are defined as in Ref.~\cite{camargo2022synchronization}.
In the model defined by Eq.~\ref{eq::Cam}, the parameter $d$ enters the savings dynamics through the nonlinear term $p x (d - y^2)$. Economically, $d$ represents the \textit{square} of the potential GDP, a threshold that separates regimes of profitable reinvestment from those of capital flight. For dimensional consistency, we define the potential GDP level as $y_{\mathrm{pot}} = \sqrt{d}$, so that the condition $y^2 < d$ corresponds to $y < y_{\mathrm{pot}}$. This formulation ensures that $y_{\mathrm{pot}}$ shares the same units as the actual GDP variable $y$, preserving interpretability within the macroeconomic context. Thus, $y_{\mathrm{pot}}$ acts as the maximum sustainable output without inflationary pressures, and the sign of $(d - y^2)$ governs the direction of savings adjustment in response to the output gap.

In this model, the parameter $m = 0.02$ denotes the variation in the marginal propensity to save, quantifying the fraction of an incremental income that is saved rather than consumed. The parameter $p = 0.4$ represents the fraction of capitalized profits, i.e., the portion of profits reinvested into the economy instead of being distributed or spent. 
In addition, $v = 0.05$ denotes the marginal propensity to consume, expressing the fraction of additional income allocated to consumption. A low value of $v$ indicate that households spend a smaller share of additional income, which may limit domestic demand but increase savings available for investment. The parameter $w = 0.1$ represents the proportion of savings; higher values of $w$ enhance the resources available for investment but may reduce consumption and, consequently, affecting overall demand. The coefficient $c = 50.0$ defines the output–capital ratio, quantifying the efficiency with which capital is converted into output. $s = 10.0$ denotes the inflow–saving ratio, measuring the extent to which domestic savings are supplemented by foreign capital inflows. Finally, the parameter $r = 0.1$ denotes the indebtedness factor, which measures the degree to which domestic investment is financed through external borrowing.

Let us start this investigation by analyzing the dynamics of the single system defined by Eq.~\ref{eq::Cam}. In this simplified setting, the system evolves independently and is solely influenced by its intrinsic parameters without any additional interactions term. 
This system admits three equilibrium points, which are obtained by solving Eq.~\ref{eq::Cam} under the assumption that $\dot{x} = 0$, $\dot{y} = 0$, and $\dot{z} = 0$. The first one, $\mathbf{E_0} = (0, 0, 0)$, is trivial. The other two $\mathbf{E_1}$ and $\mathbf{E_2}$ are symmetric and given by:
$\mathbf{E_1} =
\left(
\dfrac{r}{s}\sqrt{K},\;
\sqrt{K},\;
-\dfrac{\sqrt{K}}{c}\left(v + \dfrac{w r}{s}\right)
\right),
\qquad
\mathbf{E_2} =
\left(
-\dfrac{r}{s}\sqrt{K},\;
-\sqrt{K},\;
\dfrac{\sqrt{K}}{c}\left(v + \dfrac{w r}{s}\right)
\right),$
with $K = d + \dfrac{m s}{p r}$, for simplicity.
Assuming that all system parameters, except $d$, are fixed as previously defined, this parameter becomes the primary control variable governing the stability of the equilibrium points identified above, and implicitly the dynamical behavior of this macroeconomic model.

\begin{figure}[htp!]
\centering
\begin{tabular}{c}
\includegraphics[width=0.5\textwidth]{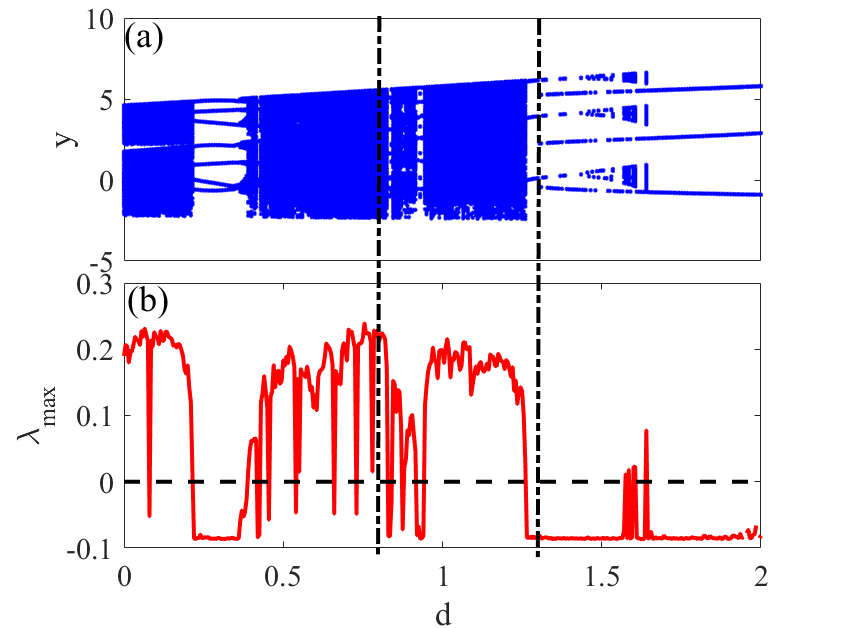}
\end{tabular}
\caption{Dynamic of a single macroeconomical system as a function of the parameter $d$: (a) Bifurcation diagram, and (b) maximum Lyapunov exponent.}
\label{fig::distd}
\end{figure}

To examine the influence of the parameter $d$ on the global dynamics of the macroeconomic model, Fig.~\ref{fig::distd} illustrates the system’s response as $d$ varies within $[0, 2]$, using two fundamental tools from dynamical systems theory: the bifurcation diagram [see Fig.~\ref{fig::distd}(a)] and the maximum Lyapunov exponent~\cite{wolf1985determining} [ see Fig.~\ref{fig::distd}(b)].
The bifurcation diagram obtained from the $y$ variable, i.e., the GDP (the other variables exhibit qualitatively similar behavior), shown in Fig.~\ref{fig::distd}(a), reveals a cascade of multi-periodic bifurcations interspersed with chaotic windows. These transitions highlight the system’s strong sensitivity to perturbation in the parameter $d$ (named potential GDP), indicating that even small changes in productive capacity can trigger qualitative shifts in the macroeconomic dynamics. The corresponding maximum Lyapunov exponent, $\lambda_{\max}$, defined in Eq.~\ref{eq::Lam} and illustrated in Fig.~\ref{fig::distd}(b), provides a quantitative characterization of these transitions, distinguishing periodic/multiperiodic $(\lambda_{\max} < 0)$ and chaotic $(\lambda_{\max} > 0)$ regimes.
\begin{equation}
\lambda_{max} = \lim_{t \to \infty} \frac{1}{t} \ln \frac{\| \delta \mathbf{X}(t) \|}{\| \delta \mathbf{X}(0) \|},
\label{eq::Lam}
\end{equation}
where $\delta \mathbf{X}(t)$ denotes an infinitesimal perturbation of the trajectory of the system at time $t$. Let us remember that, the maximum Lyapunov exponent quantifies the average exponential rate of divergence between nearby trajectories in phase space~\cite{wolf1985determining}. From an economic perspective, it also measures the sensitivity of the system to initial conditions: when $\lambda_{\max}>0$, small perturbations in savings, production, or capital inflows grow over time, leading to unpredictable macroeconomic fluctuations, whereas $\lambda_{\max}<0$ indicates convergence toward stable economic trajectories for any kind of perturbation.

\section{Network of coupled macroeconomical systems}\label{sec::sec3}
\subsection{Fitness parameter and network topology}\label{sec::subsec31}

\begin{figure*}[htp!]
\centering
\begin{tabular}{c}
\hspace{0.0 cm}
\includegraphics[width=0.950\textwidth]{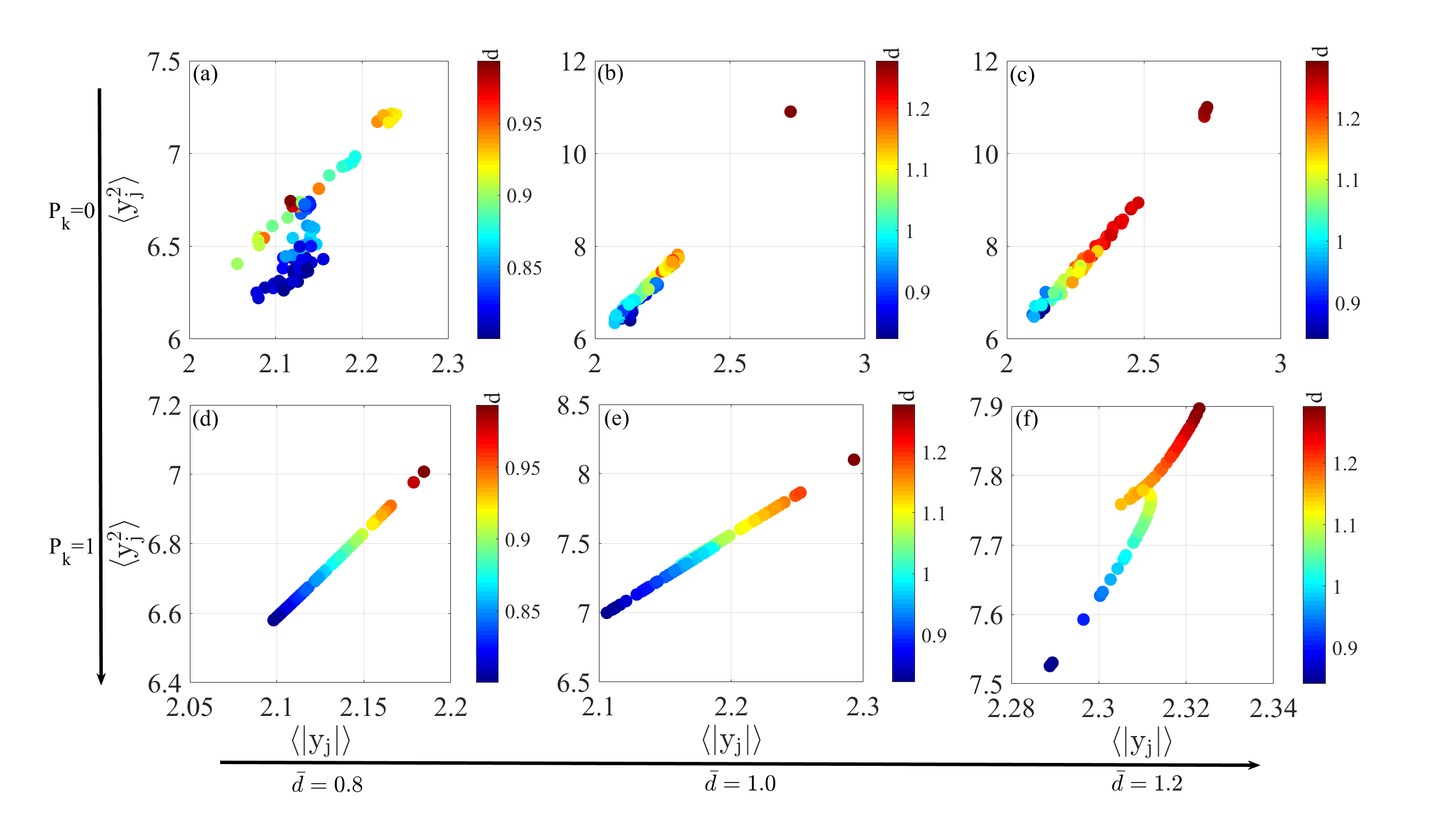}
\end{tabular}
\caption{Relationship between the time-averaged squared GDP, $\langle y_j^2 \rangle$, and the time-averaged absolute GDP, $\langle |y_j| \rangle$, for all nodes. The parameter $d$ is sampled from a log-normal distribution over the interval $[0.8, 1.3]$, with $\ln d \sim \mathcal{N}(\bar d, \chi^2)$ and $\chi = 0.1$. The first row (a–c) represents a fully disconnected network, where the degree distribution is zero ($P_k = 0$), while the second row (d–f) corresponds to a fully connected network, with $P_k = 1$. From left to right, the columns correspond to (a,d) $\bar d = 0.8$, (b,e) $\bar d = 1.0$, and (c,f) $\bar d = 1.2$. The color scale indicates the values of $d$.
}
\label{fig::Cor_dy}
\end{figure*}

We consider a network composed of $N$ macroeconomic units, indexed by $j=1,\ldots,N$, each governed by the model (see Eq.~\ref{eq::Cam}) introduced in Sec.~\ref{sec::sec2}. While the structural equations are identical across nodes, their dynamics may differ through the parameter $d_j$ associated to each node, which captures variations in productive capacity among economies. The values $d_j$ are sampled from a log-normal distribution over $[d_{\min}, d_{\max}]$, with $\ln d \sim \mathcal{N}(\bar d, \chi^2)$. $\bar d$ and $\chi$ denote the mean and standard deviation of the associated distribution, respectively. The resulting system, defined analytically by Eq.~\ref{eq::Cam_het}, thus represents a heterogeneous network of interacting economies.

\begin{equation}
    \begin{cases}
        \dot{x_j} = my_j + px_j(d_j - y_j^2), \\
        \dot{y_j} = vy_j + wx_j + cz_j, \\
        \dot{z_j} = sx_j - ry_j + \frac{\sigma}{k_j} \sum_{k=1}^N A_{jk} (z_k - z_j).
    \end{cases}
    \label{eq::Cam_het}
\end{equation}
where $\sigma$ and $k_j$ are respectively the coupling strength and the degree associated to the node $j$.
The economies interact through a network structure that captures flows of trade, investment, or financial influence among them. Analytically, these interactions are encoded in the adjacency matrix $A=\{a_{jk}\}$ defined in Eq.~\ref{eq::A_mat}  and scaled by a coupling coefficient $\sigma$, which regulates the degree of interdependence between nodes $k$.

\begin{equation}
    A_{jk} = 
    \begin{cases}
        1  \quad \text{with} \quad   P_{jk}(d_j,d_k) \quad \text{($j \neq k$)}\\
        0  \quad \text{with} \quad   1 - P_{jk}(d_j,d_k).
    \end{cases}
    \label{eq::A_mat}
\end{equation}
The term $P_{jk}(d_j,d_k)$ defines the probability that two economies are connected. As in real economic models, connections and exchanges are closely related to the GDP. To illustrate this point, Fig.~\ref{fig::Cor_dy} shows the relationship between $\langle y_j^2 \rangle$ (the time average of $y_j^2$ over a sufficiently long time 
window $T$) and $\langle |y_j| \rangle$ (the time average of $|y_j|$ over the same 
interval), for two extreme network topologies: \\
(i) the first corresponds to an uncoupled 
network in which all nodes evolve independently, i.e., with a null degree 
distribution $P_k=0$ (see the first row of Fig.~\ref{fig::Cor_dy}(a--c)); \\
(ii) the second case is a fully connected network, characterized by a degree distribution $P_k=1$ 
(see the second row of Fig.~\ref{fig::Cor_dy}(d--f)).
In both all cases, a clear 
correlation between $\langle y_j^2 \rangle$ and $\langle |y_j| \rangle$ is observed, 
independently of the value of $d$.

Despite this correlation, $\langle |y_j| \rangle$ and $\langle y_j^2 \rangle$ remain dependent on the 
instantaneous state of the system, the initial conditions, the coupling strength, 
and endogenous fluctuations. By contrast, the potential GDP $d_j$, which quantifies the 
productive potential of an economy and of which $y_j(t)$ is only a dynamical 
manifestation, acts as a fundamental control parameter of the local dynamics through 
the nonlinear term $p x_j (d_j - y_j^2)$ (see Fig.~\ref{fig::distd}). A dimensional analysis further shows that $[d]=[y^2]$, 
implying that $d_j$ directly controls the order of magnitude of $\langle y_j^2 \rangle$, and thus indirectly that of $\langle |y_j| \rangle$. 
Consequently, defining the probability of connection between two nodes $j$ and $k$ based on their potential GDPs $d_j$ and $d_k$ leads to a stable and economically interpretable network topology, independent of short-term fluctuations.

Therefore, the term $P_{jk}(d_j,d_k)$ appearing in Eq.~\ref{eq::A_mat} is defined by the fitness function~\cite{garlaschelli2004fitness} introduced in Eq.~\ref{eq::p_con}, and specifies the probability that two nodes or economies $j$ and $k$ are connected, depending on their respective parameters $d_j$ and $d_k$. 
\begin{equation}
    P_{jk}(d_j,d_k) = \frac{\delta d_j d_k}{1+\delta d_j d_k},
    \label{eq::p_con}
\end{equation}
with $\delta$ the fitness parameter, which can be interpreted as a proxy for commercial and financial integration, as it modulates the density of links between economic entities.
By adjusting this parameter $\delta$, the network can evolve from an isolated state $(\delta \to 0)$ to a fully interconnected economy $(\delta \to +\infty)$,  passing through  intermediate sparse configurations. This gradual change in the network structure embodies the heterogeneity of global economies, where disparities in productive capacity and capital flows shape both structure and dynamics.

For the set of $N$ interconnected macroeconomic systems defined by Eq.~\ref{eq::Cam_het}, we assume that the parameter $d$ of each node are drawn from a log-normal distribution truncated to the interval $[0.8,\,1.3]$, bounded by the dashed vertical lines in Fig.~\ref{fig::distd}, with a mean value $\bar{d}=1$. Within this range of $d$ values, the systems without any interaction exhibits both multi-periodic and chaotic attractors, as shown by the bifurcation diagram and the corresponding maximum Lyapunov exponent in Fig.~\ref{fig::distd}. This distribution introduces controlled heterogeneity among the nodes of the network, representing economies with varying productive capacities. 
To illustrate the structural properties of the network generated from this distribution, we define the diffusion matrix in its Laplacian form, $L = D - A$, where $D$ is the degree matrix and $A$ the adjacency matrix as defined in Eq.~\ref{eq::A_mat}. From the dynamical systems perspective, this Laplacian defines how information propagates between nodes of the network and determines the collective behavior of the interconnected systems. It is well known that the eigenvalues of $L$ describe fundamental structural properties: the first and smallest eigenvalue, $\lambda_1 = 0$, corresponds to the homogeneous mode of the network, while the second smallest eigenvalue, $\lambda_2$, known as the \textit{Fiedler value}~\cite{fiedler1973algebraic} measures the overall connectivity. From an economic perspective, the Laplacian matrix captures the pattern of interdependence among economies, its elements quantify how changes in one economy diffuse through investment, trade, or financial channels to others. Therefore, a larger $\lambda_2$ indicates a more cohesive network, where shocks or fluctuations are rapidly transmitted and absorbed, whereas smaller $\lambda_2$ values reflect weaker connectivity and a higher tendency toward fragmentation or asynchronous behavior among economies.

\begin{figure}[htp!]
\centering
\begin{tabular}{c}
\includegraphics[width=0.45\textwidth]{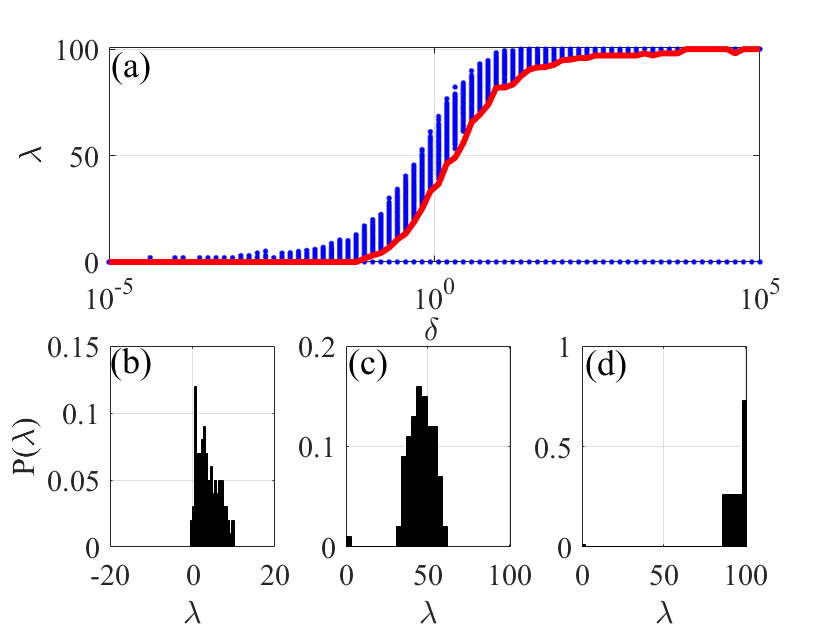}\\
\includegraphics[width=0.45\textwidth]{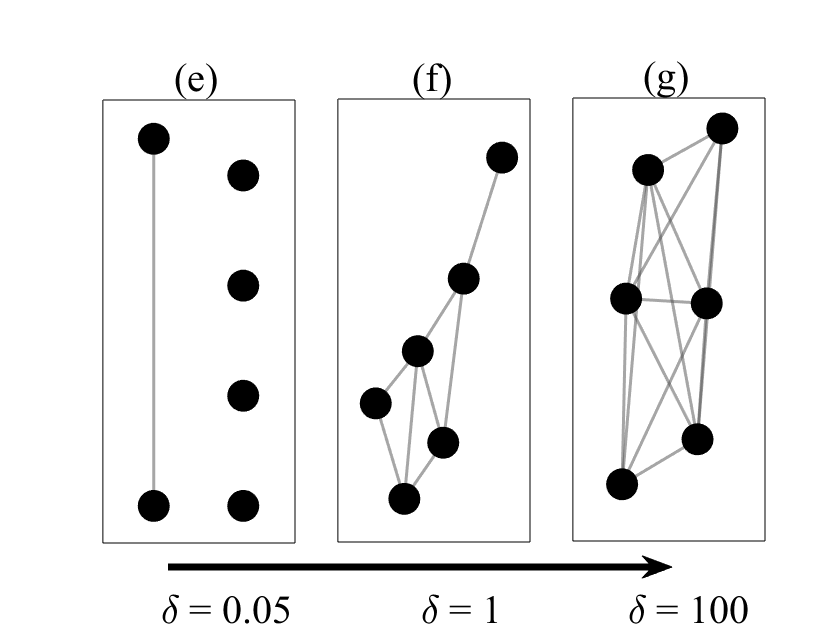}
\end{tabular}
\caption{Evolution of the network topology with the fitness parameter $\delta$. (a) Semi-logarithmic plot of the Laplacian spectrum for a network of $N=100$ nodes. Blue dots denote all eigenvalues, and red line mark the Fiedler value $\lambda_2$. (b–d) Eigenvalue distributions for $\delta=5\times10^{-2}$, $1$, and $10^2$, respectively. (e–g) Corresponding network configurations for a smaller system of $6$ nodes (just for illustration), illustrating the transition from a fragmented and isolated network/economic structure to a highly integrated one as $\delta$ increases, passing through sparse network for intermediate $\delta$.}
\label{fig::op_e}
\end{figure}

Fig.~\ref{fig::op_e} illustrates the influence of the fitness parameter $\delta$ on the network topology. As mentioned previously, $\delta$ modulates the density of the link and, consequently, the spectral properties of the Laplacian matrix. This effect is clearly visible in Fig.~\ref{fig::op_e}(a), where the  spectrum undergoes a significant variation: for $\delta \to 0$, all eigenvalues approach zero, corresponding to an unconnected graph that represents a fragmented global economy with limited trade interactions; conversely, for $\delta \to +\infty$, the eigenvalues converge towards $\lambda_j \to N$ (except $\lambda_1$ which is zero), indicating a fully connected structure that mirrors a highly integrated economic system with intense exchanges among countries. Fiedler values, highlighted in red, increase from zero in the unconnected case to $N$ for a fully connected network, reflecting the progressive strengthening of structural and economic cohesion~\cite{jackson2003strategic}. Figs.~\ref{fig::op_e}(b--d) show the corresponding eigenvalue distributions for $\delta = \{5\times10^{-2},\,1,\,10^2\}$. As $\delta$ increases, the network transitions between two limiting regimes: a fragmented or weakly connected structure for small $\delta$ [Fig.~\ref{fig::op_e}(e)], a partially integrated configuration at intermediate $\delta$ [Fig.~\ref{fig::op_e}(f)], and an almost fully connected topology for large $\delta$ [Fig.~\ref{fig::op_e}(g)], indicative of a globalized economic system with high connectivity and interdependence.

\subsection{Homogeneous network with identical potential GDP}\label{sec::subsec3B}

We consider in this subsection an idealized homogeneous network in which all nodes are identical, corresponding to a fully symmetric economic system where every system/node has the same parameter, i.e., $d_j = d$ for all $j = 1,\dots,N$. This network is governed by the following Eq.~\ref{eq::homomf}:

\begin{equation}
    \begin{cases}
        \dot{x_j} = my_j + px_j(d - y_j^2), \\
        \dot{y_j} = vy_j + wx_j + cz_j, \\
        \dot{z_j} = sx_j - ry_j + \frac{\sigma}{k_j} \sum_{k=1}^N A_{jk} (z_k - z_j).
    \end{cases}
    \label{eq::homomf}
\end{equation}
In this homogeneous configuration, the connection probability $P_{jk}(d_j,d_k)$ defined by Eq.~\ref{eq::p_con} takes the simplified form defined in Eq.~\ref{eq::homo_p_conmf}.
\begin{equation}
    P_{jk}(d) = \frac{\delta d^2}{1+\delta d^2},
    \label{eq::homo_p_conmf}
\end{equation}

\begin{figure*}[htp!]
\centering
\begin{tabular}{c}
\hspace{0.0 cm}
\includegraphics[width=0.950\textwidth]{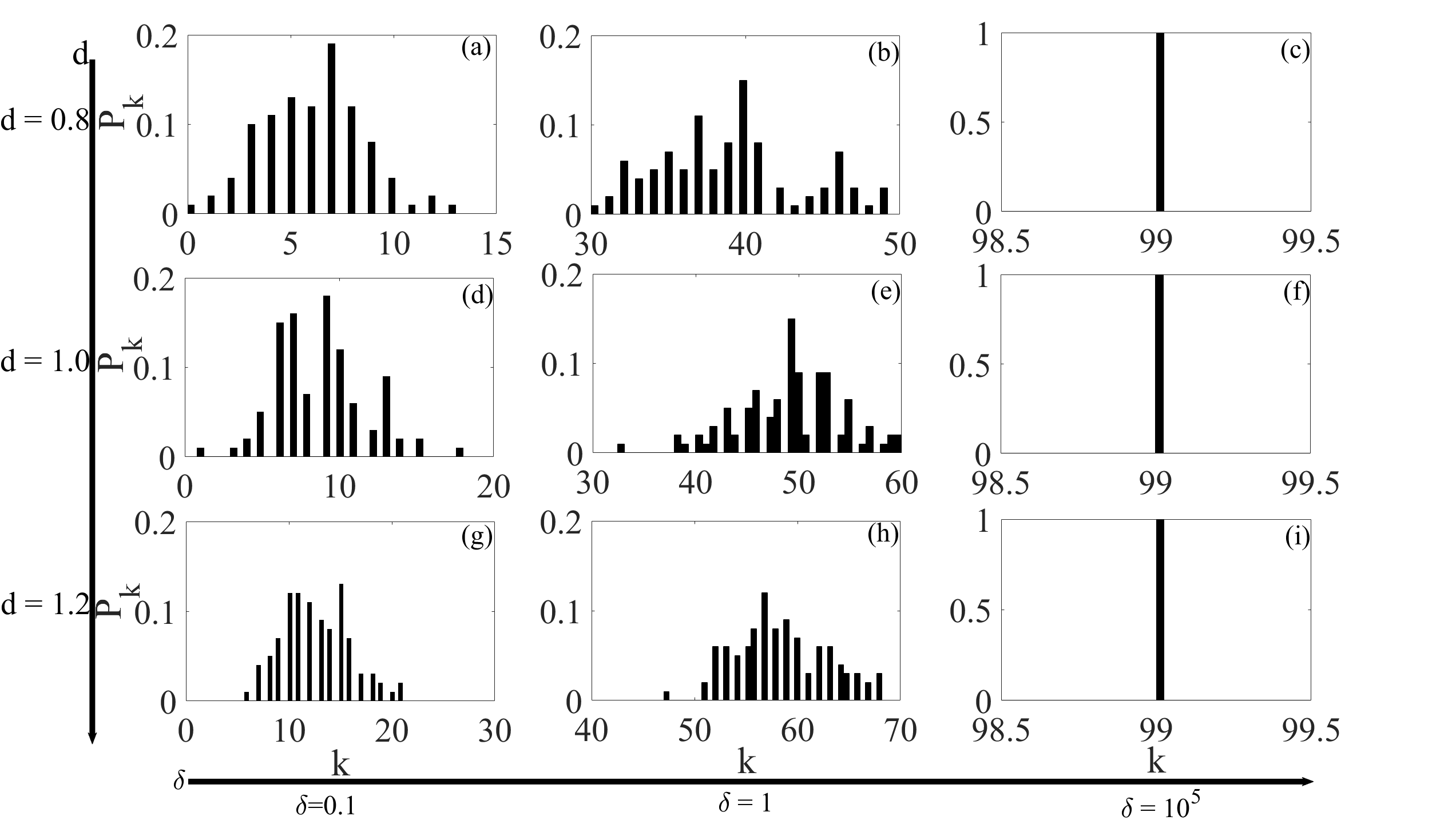}
\end{tabular}
\caption{Degree distribution of the network for three values of the potential GDP, $d \in \{0.8,\,1.0,\,1.2\}$, and three fitness parameters, $\delta \in \{10^{-1},\,1.0,\,10^{5}\}$, illustrating how both parameters shape the network connectivity.}
\label{fig::disthet}
\end{figure*}

To illustrate the impact of this probability on the network structure, Fig.~\ref{fig::disthet} shows the degree distribution of the network for three values of the parameter $d = \{0.8,\,1.0,\,1.3\}$, representing in an artificial way, economies of low, medium, and high productivity, and for three fitness parameters, $\delta = \{10^{-1},\,1.0,\,10^{5}\}$, which modulate the network’s connectivity and then the topology.
Regardless of the parameter $d$, the fitness parameter $\delta$ governs the network topology in a similar manner, leading to equivalent topologies. For large values of $\delta$ (e.g., $\delta = 10^5$), the network becomes fully connected, and the degree of each node approaches $k_j \approx N-1$, as shown in panels~\ref{fig::disthet}(c,f,i), while topological fluctuations become negligible. The average degree follows directly from $\langle k \rangle \approx (N-1)P(d)$, yielding
\begin{equation}
    \langle k \rangle \approx (N-1)\frac{\delta d^2}{1+\delta d^2}.
    \label{eq::degm}
\end{equation}
This expression shows that the average degree increases with both $\delta$ and $d$, saturating at $\langle k \rangle \approx N-1$ when $\delta  \to +\infty$. From a structural viewpoint, this behavior reflects the smooth transition from sparse to fully connected topologies as $\delta$ grows. From economic perspective, higher values of $\delta$ or $d$ correspond to stronger trade or investment capacities, leading to denser interconnections among economies and a higher level of global integration.

\subsubsection{Homogeneous mean-field (HMF) approximation: $\delta \to +\infty$}\label{sec::subsubsec3B1}
To gain analytical insight into the collective dynamics of the system, we first examine the idealized limit of a fully connected and homogeneous network. We focus here on the case $\delta \to +\infty$ and $d_j = d$ for all $j$. This configuration allows for a mean-field approximation using the \textit{Homogeneous Mean-Field Theory}~\cite{montbrio2020exact,mirollo1990amplitude}, which captures the aggregate behavior of the system under conditions of perfect economic integration (i.e., fully connected network configuration). Under this HMF approximation, each node interacts with the mean behavior of the entire network rather than with individual nodes.

For simplicity, we focus on the third variable, $z_j$, in Eq.~\ref{eq::homomf}, which includes the coupling function and explicitly encodes the network topology, as reproduced in Eq.~\ref{eq::eqz}.
\begin{equation}
    \dot{z}_j = s x_j - r y_j + \frac{\sigma}{k_j}\sum_{k=1}^N A_{jk}(z_k - z_j),
    \label{eq::eqz}
\end{equation}
where the degree of node $j$ is $k_j = \sum_{k=1}^N A_{jk}$.
The diffusive coupling term can be rewritten as:
\begin{equation}
    \dot{z}_j = s x_j - r y_j + \sigma\left(\frac{1}{k_j}\sum_{k=1}^N A_{jk}z_k - z_j\right),
\end{equation}
Under the homogeneous mean-field approximation,
the adjacency matrix $A_{jk}$ is replaced by its expectation value $\mathbb{E}[A_{jk}]$:
\begin{equation}
    A_{jk} \longrightarrow \mathbb{E}[A_{jk}] = P_{jk}(d),
\end{equation}
where $\mathbb{E}$ is the esperance. Then, Eq.~\ref{eq::eqz} could be rewritten as:
\begin{equation}
    \dot{z}_j = s x_j - r y_j + \sigma\left(\frac{1}{k_j}\sum_{k=1}^N P_{jk}(d)z_k - z_j\right),
    \label{eq::eqpd}
\end{equation}
Let us remember that the connection probability is uniform, i.e. $P_{jk}(d) \equiv \frac{\delta d^2}{1+\delta d^2}$, as we consider the same value of $d$ for all nodes and then, Eq.~\ref{eq::eqpd} becomes:
\begin{equation}
    \dot{z}_j = s x_j - r y_j + \sigma\left(\frac{1}{(N-1)}\sum_{k=1}^N \frac{\delta d^2}{1+\delta d^2} z_k - z_j\right),
\end{equation}
For large networks, we use the approximation $(N-1)\simeq N$, which means that each node interacts with an increasingly large population and the coupling can be expressed through a global average. This naturally leads to a homogeneous mean-field description, where the influence of the network on node $j$ is captured by the difference between the global mean $\bar{z} = \frac{1}{N} \sum_{k=1}^N z_k$ and the local value $z_j$. Hence, we obtain the standard homogeneous mean-field form:

\begin{equation}
    \begin{cases}
        \dot{x}_j = my_j + px_j(d - y_j^2), \\
        \dot{y}_j = vy_j + wx_j + cz_j, \\
        \dot{z}_j = sx_j - ry_j + \sigma (\bar{z} - z_j),
    \end{cases}
    \label{eq::homomff}
\end{equation}

\begin{figure}[htp!]
\centering
\begin{tabular}{c}
\hspace{-0.80 cm}
\includegraphics[width=0.570\textwidth]{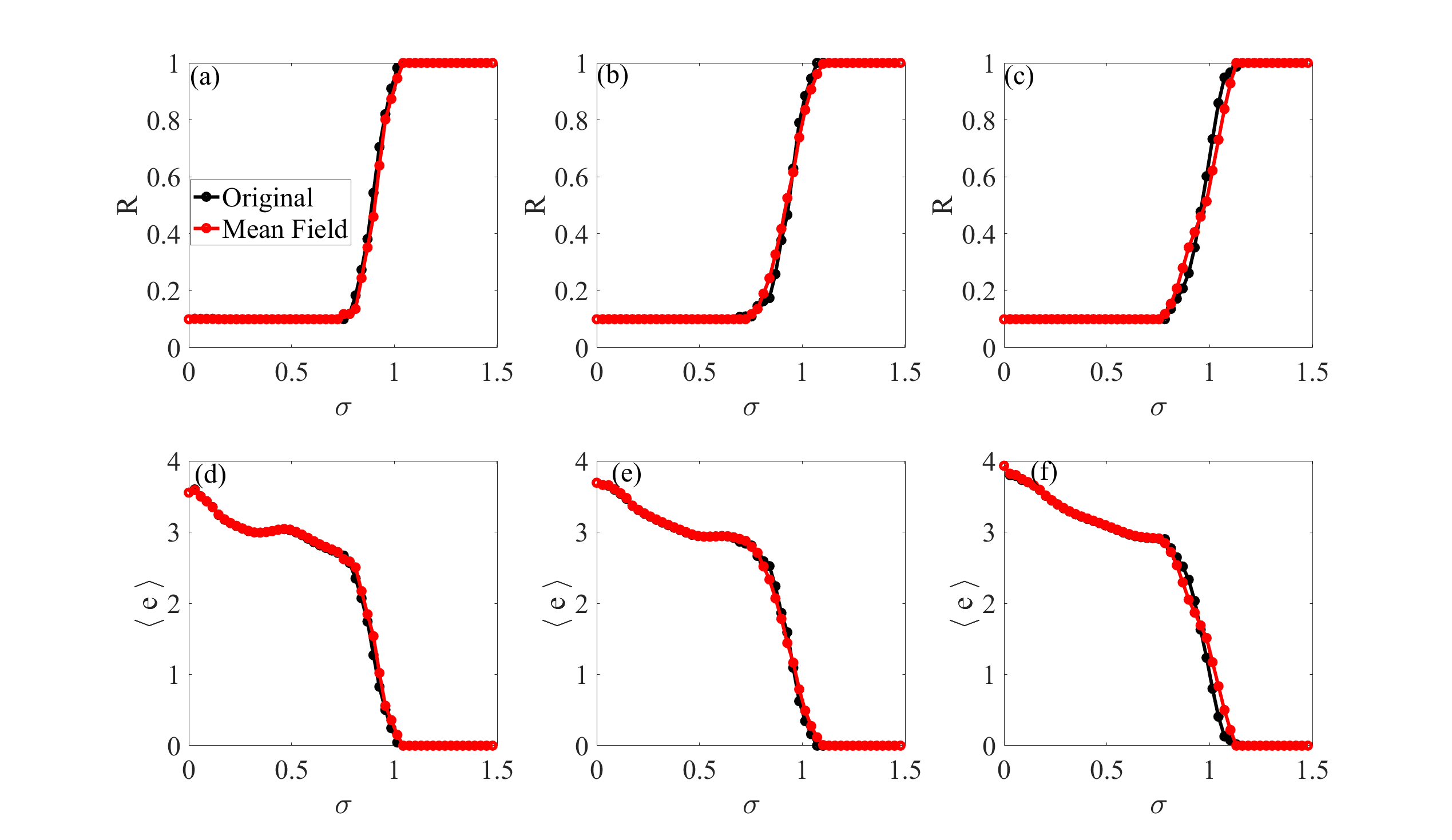}
\end{tabular}
\caption{Synchronization transitions for a homogeneous network (nodes and topology) as a function of the coupling strength $\sigma$, for different values of the parameter $d$. The first row (panels ($a$–$c$)) shows the average order parameter, while the second row (panels ($d$–$f$)) shows the average synchronization error. From left to right, the first (panels ($a$, $d$)), second (panels ($b$, $e$)), and third (panels ($c$, $f$)) columns correspond to $d = 0.8$, $d = 1$, and $d = 1.2$, respectively, $\delta=10^5$.}
\label{fig::MER_full}
\end{figure}

Let us numerically integrate both the original model, Eq.~\ref{eq::homomf}, and its mean-field reduction, Eq.~\ref{eq::homomff}, using a fourth-order Runge--Kutta algorithm with a time step $dt=0.05$ for $n=5\times10^4$ iterations to investigate the collective dynamics of the macroeconomic network. The solutions are analyzed after discarding the initial $75\%$ of the time series as a transient phase. Fig~\ref{fig::MER_full} shows the effect of the coupling strength $\sigma$ on the two descriptions; the original model (black) and the mean-field approximation (red) averaged over 20 different realizations for each value of $\sigma$. Panels~\ref{fig::MER_full}(a--c) report the average order parameter $R$ (defined in Eq.~\ref{eq::opp} from the variable $y$) as a function of $\sigma$ for three values of $d \in \{0.8,\,1.0,\,1.2\}$, respectively. The results indicate that the collective dynamics are essentially insensitive to $d$: for all three values, the onset of synchronization is nearly identical and is captured equally well by the original and mean-field models. We thus conclude that, for a fully connected network of identical economies, the parameter $d$ does not affect the emergence of collective synchronization—economic disparities are neutralized under homogeneous, fully integrated conditions.

Let us define the concept of the order parameter, originally introduced by Kuramoto and Battogtokh~\cite{kuramoto2002coexistence}, which provides a standard measure of phase synchronization in systems of coupled oscillators. Its computation relies on extracting the instantaneous phase from each individual time series. For a time-dependent signal $s(\tau)$, the phase is obtained from the analytic signal constructed via the Hilbert transform $\tilde{s}(\tau)$, namely
\begin{equation*}\label{eq::ht}
\psi(\tau) = s(\tau) + i\,\tilde{s}(\tau) = r(\tau) e^{i\varphi(\tau)},
\end{equation*}
where $i^2=-1$, $r(\tau)$ is the instantaneous amplitude, and $\varphi(\tau)$ the instantaneous phase of $s(\tau)$. 
For each oscillator $j$, the instantaneous phase $\varphi_j(\tau)$ is then given by
\begin{equation*}\label{eq::pha}
\varphi_j(\tau) = \tan^{-1}\!\left(\frac{\tilde{s}_j(\tau)}{s_j(\tau)}\right).
\end{equation*}
The global phase coherence of a network with $N$ oscillators is quantified by the complex order parameter
\begin{equation}\label{eq::op}
r = \left| \frac{1}{N} \sum_{j=1}^N e^{i\varphi_j} \right|.
\end{equation}
In our macroeconomic framework, $r$ captures the degree of collective alignment among economies, that is, how synchronized their cyclical behaviors are over time. For the $M$ numerical realizations performed for each control parameter, the average order parameter shown in Fig.~\ref{fig::MER_full} is computed as
\begin{equation}\label{eq::opp}
R = \frac{1}{M} \sum_{k=1}^M r_k,
\end{equation}
with the same properties as the original Kuramoto parameter: $R$ ranges from $0$ (complete desynchronization) to $1$ (full phase synchronization across the network). High values of $R$ therefore indicate strong macroeconomic coherence, while low values correspond to asynchronous or weakly correlated economic activity.
Therefore, the \textit{GDP phase synchronization}~\cite{rosenblum1996phase} shown in Fig.~\ref{fig::MER_full} refers to the situation in which the economies of several countries experience fluctuations—periods of growth and recession—that are temporally aligned. In other words, their business cycles oscillate in phase, even if the amplitudes of these fluctuations differ. Such temporal coordination of economic activity reflects the degree of coherence among interconnected economies. This phenomenon is of particular relevance for understanding global interdependencies, especially within economically integrated regions such as the European Union. A high level of GDP cycle synchronization, as emphasized by Imbs~\cite{imbs2004trade}, typically signals strong trade and financial integration, but it may also heighten systemic vulnerability by amplifying the transmission of economic shocks across the global network.

Panels~\ref{fig::MER_full}(d--f) show the average synchronization error $\langle e \rangle$ (defined in Eq.~\ref{eq::errf}) as a function of the coupling strength $\sigma$ for the same parameter values $d$ used in Panels~\ref{fig::MER_full}(a--c). The results of the two models are nearly indistinguishable, consistent with those obtained for the order parameter (see panels~\ref{fig::MER_full}(a--c)), confirming the validity of the mean-field approximation in capturing the collective behavior of the system. A decrease in $\langle e \rangle$ (resp. an increase in $R$) with increasing $\sigma$ (which can be interpreted as a rise in mutual confidence or trust among economies) reflects the emergence of a fully synchronization~\cite{pikovsky2001synchronization} (resp. phase synchronization) leading to synchronized macroeconomic cycles, where stronger coupling mimics intensified trade or greater openness to exchange.

For a single simulation, the instantaneous synchronization error computed over the time window $[t,\,T]$ is given by
\begin{equation}
e = \frac{1}{T} \int_{t}^{T+t} 
\left( \frac{1}{N(N-1)} \sum_{j,k=1}^N \|\mathbf{X}_j(t) - \mathbf{X}_k(t)\|^2 \right)^{1/2} dt,
\label{eq::err}
\end{equation}
and, for $M$ independent realizations, the average synchronization error is
\begin{equation}
\langle e \rangle = \left| \frac{1}{M} \sum_{k=1}^M e_k \right|,
\label{eq::errf}
\end{equation}
where $\|\cdot\|$ denotes the Euclidean distance between the states $\mathbf{X}_j=(x_j, y_j, z_j)$ and $\mathbf{X}_k=(x_k, y_k, z_k)$ at time $t$.

\begin{figure*}[htp!]
\centering
\begin{tabular}{ccc}
\includegraphics[width=0.3\textwidth]{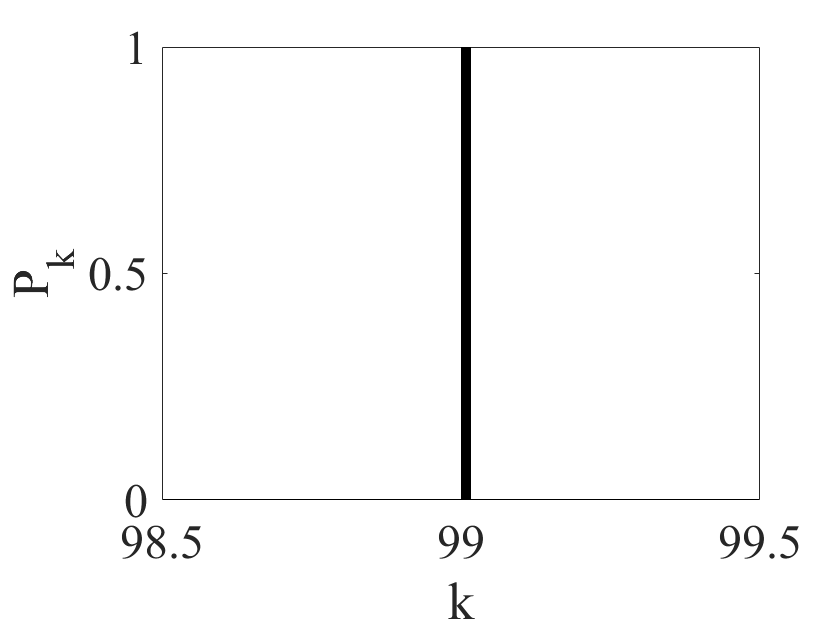}&
\includegraphics[width=0.3\textwidth]{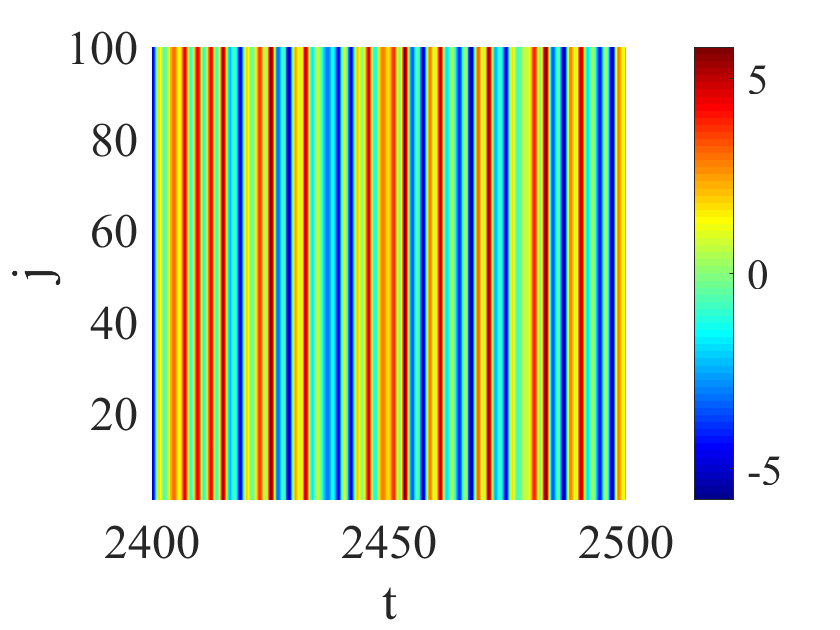}&
\includegraphics[width=0.3\textwidth]{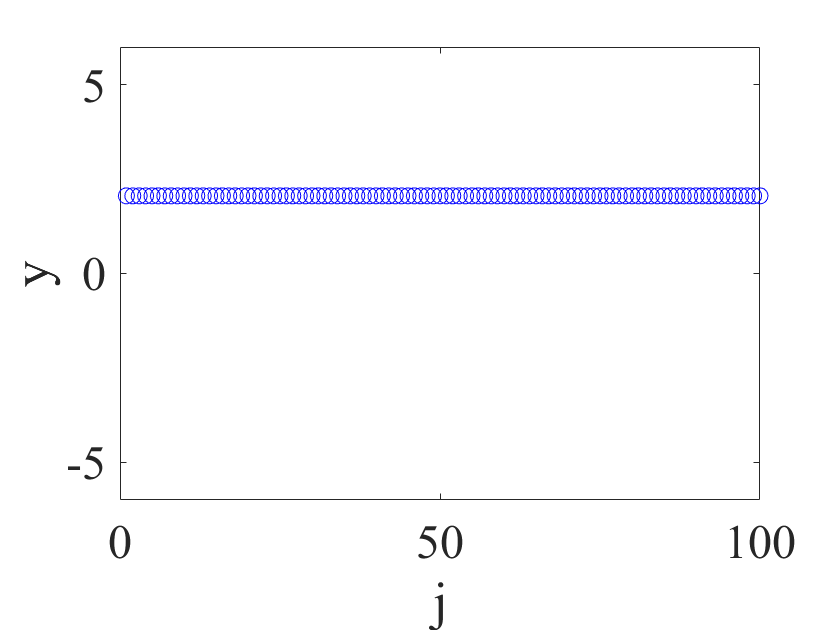}\\
(a) & (b) & (c)
\end{tabular}
\caption{Network degree distribution and collective dynamics in a network of $100$ identical oscillators. (a) Degree distribution confirming full connectivity, (b) representative time series showing complete synchronization, and (c) snapshot of the GDP variable $y$ at $t = 2500$ for $d = 1$, $\sigma = 1.1$, and $\delta = 10^5$.}
\label{fig::MER_full_port}
\end{figure*}
To illustrate the synchronization pattern discussed previously, we focus on Fig.~\ref{fig::MER_full}(b) for the case of a $d = 1$ in a fully connected configuration ($\delta = 10^5$) and a coupling strength $\sigma = 1.1$. Fig.~\ref{fig::MER_full_port} shows the degree distribution in panel~\ref{fig::MER_full_port}(a), confirming that large values of $\delta$ lead to an (almost) fully connected network. In this regime, the average degree approaches its maximum value, i.e., $k \to (N-1)$. Panel~\ref{fig::MER_full_port}(b) displays the representative time series exhibiting perfect alignment among the oscillators, i.e., complete synchronization; and panel~\ref{fig::MER_full_port}(c) illustrate a snapshot of the variable $y$ for all nodes at $t = 2500$ (the other variables $x$ and $z$ exhibit the same behavior), where all economies evolve coherently with identical amplitude. This behavior represents an idealized regime of full  coordination, in which all economic units share synchronized growth and contraction phases, analogous to phase-locking~\cite{ikeda2013synchronization} in coupled oscillator systems. While such coherence implies maximum stability and predictability, it also reflects the loss of individual dynamical diversity, making the system potentially more vulnerable to collective shocks.

\subsubsection{Limit of the homogeneous mean-field approximation}\label{sec::subsubsec3B2}

The objective of this subsection is to examine the limits of the homogeneous mean-field (HMF) model presented in Sec.~\ref{sec::subsec3B}. To this end, we consider a heterogeneous network structure by adjusting the fitness parameter $\delta$, which controls the connection density and thus modulates the network topology. In contrast to the homogeneous case, where the HMF model assumes that all nodes are statistically equivalent and interact through an average global field, this simplification may overlook heterogeneities in node properties, such as degree distributions or local dynamical behavior. To highlight these limitations, we perform a comparative study of the same model analyzed in Sec.~\ref{sec::subsubsec3B1}, but with a heterogeneous topology obtained for $\delta = 1$. The corresponding degree distributions, illustrated in Fig.~\ref{fig::disthet}(b,e,h), are shown for three values of $d = 0.8$, $1.0$, and $1.2$. For each configuration, the degree distribution follows a pattern close to a Poisson law, typical of random networks. From an economical perspective, this heterogeneity represents the coexistence of economies with different levels of connectivity and performance. It enables a deeper exploration of how structural and dynamical diversity influence collective coordination and macroeconomic stability within the network.

\begin{figure}[htp!]
\centering
\begin{tabular}{c}
\hspace{-0.90 cm}
\includegraphics[width=0.58\textwidth]{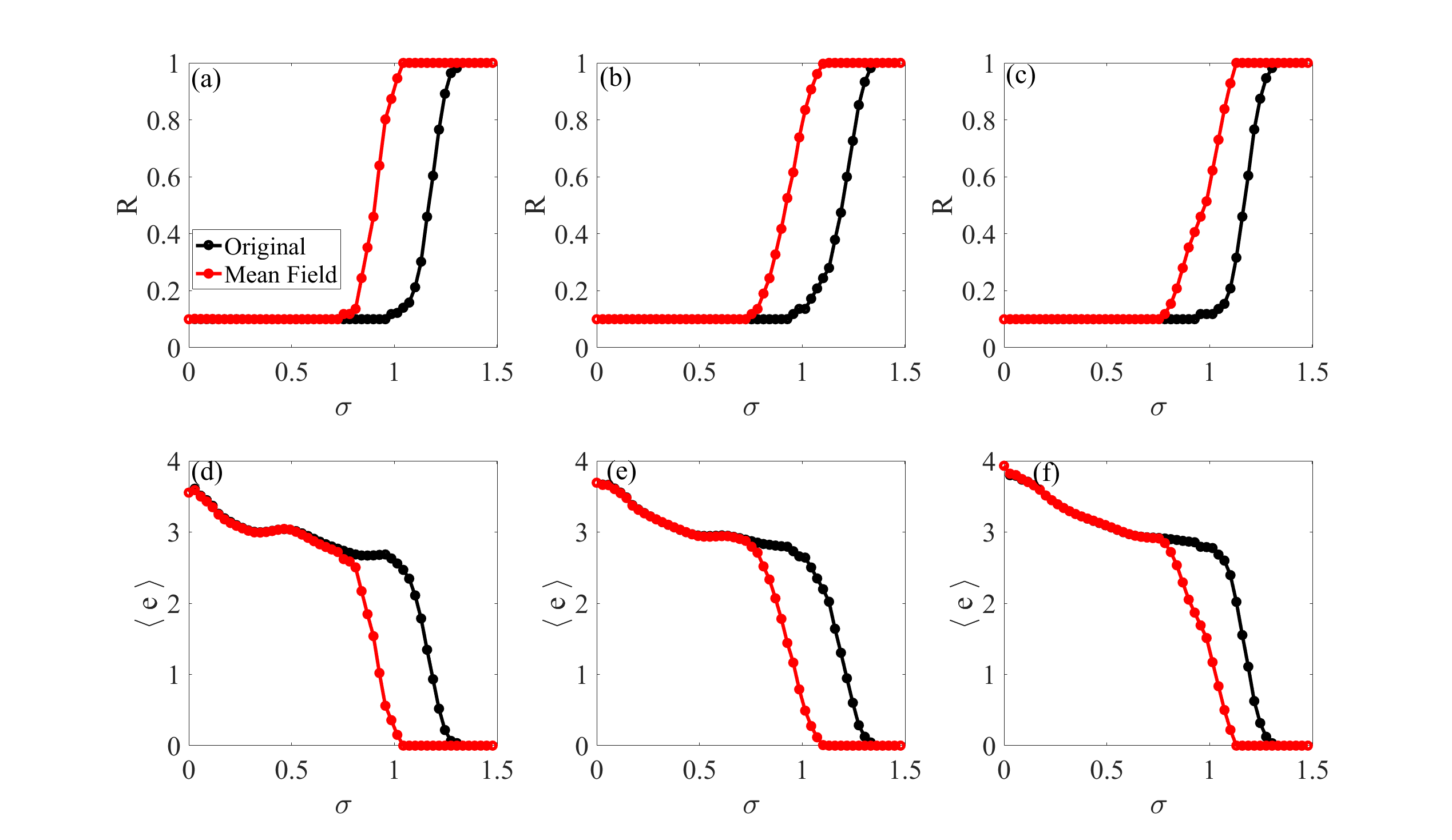}
\end{tabular}
\caption{Synchronization transitions in a heterogeneous network with $\delta = 1$ and identical oscillators as a function of the coupling strength~$\sigma$. The top row (a--c) shows the average order parameter, while the bottom row (d--f) shows the average synchronization error. Columns from left to right correspond to $d = 0.8$, $1.0$, and $1.2$, respectively.}
\label{fig::ER_full}
\end{figure}

Fig.~\ref{fig::ER_full} presents the average order parameter and the average synchronization error obtained from 20 independent realizations of the model defined by Eq.~\ref{eq::homomff} under the homogeneous mean-field (HMF) approximation but on heterogeneous network topology (with the degree distribution following a poisson law). These results illustrate the emergence of synchronization in this complex network of interacting dynamical systems. In the top row (panels~\ref{fig::ER_full}(a--c)), the average order parameter $R$ increases sharply with the coupling strength $\sigma$, marking a transition from a disordered, incoherent regime to a coherent, synchronized state. This collective transition reflects the onset of large-scale coordination among the economic systems. A quick comparison between the original model (black curves) and the mean-field approximation model (red curves) reveals a systematic shift: the mean-field model predicts an earlier onset of synchronization, corresponding to a lower critical coupling. In the bottom row (panels~\ref{fig::ER_full}(d--f)), the average synchronization error $\langle e \rangle$ decreases correspondingly with increasing $\sigma$, vanishing in the fully synchronized regime. Again, the mean-field approximation underestimates the critical coupling value, anticipating the suppression of desynchronization. \\
The systematic difference observed between the original model and the mean-field approximation model arises from the neglect of structural heterogeneity in the latter. In the original model, each node interacts according to the actual network topology, where variations in degree and local connectivity introduce non-uniform coupling strengths across the system. These differences slow down the collective alignment process, making global synchronization more difficult to achieve. In contrast, the homogeneous mean-field approach replaces all local interactions with an averaged global field, effectively smoothing out fluctuations and ignoring the influence of highly connected or peripheral nodes. As a result, the HMF model systematically anticipates the onset of synchronization, predicting a lower critical coupling strength.

\begin{figure}[htp!]
\centering
\begin{tabular}{c}
\hspace{-0.50 cm}
\includegraphics[width=0.55\textwidth]{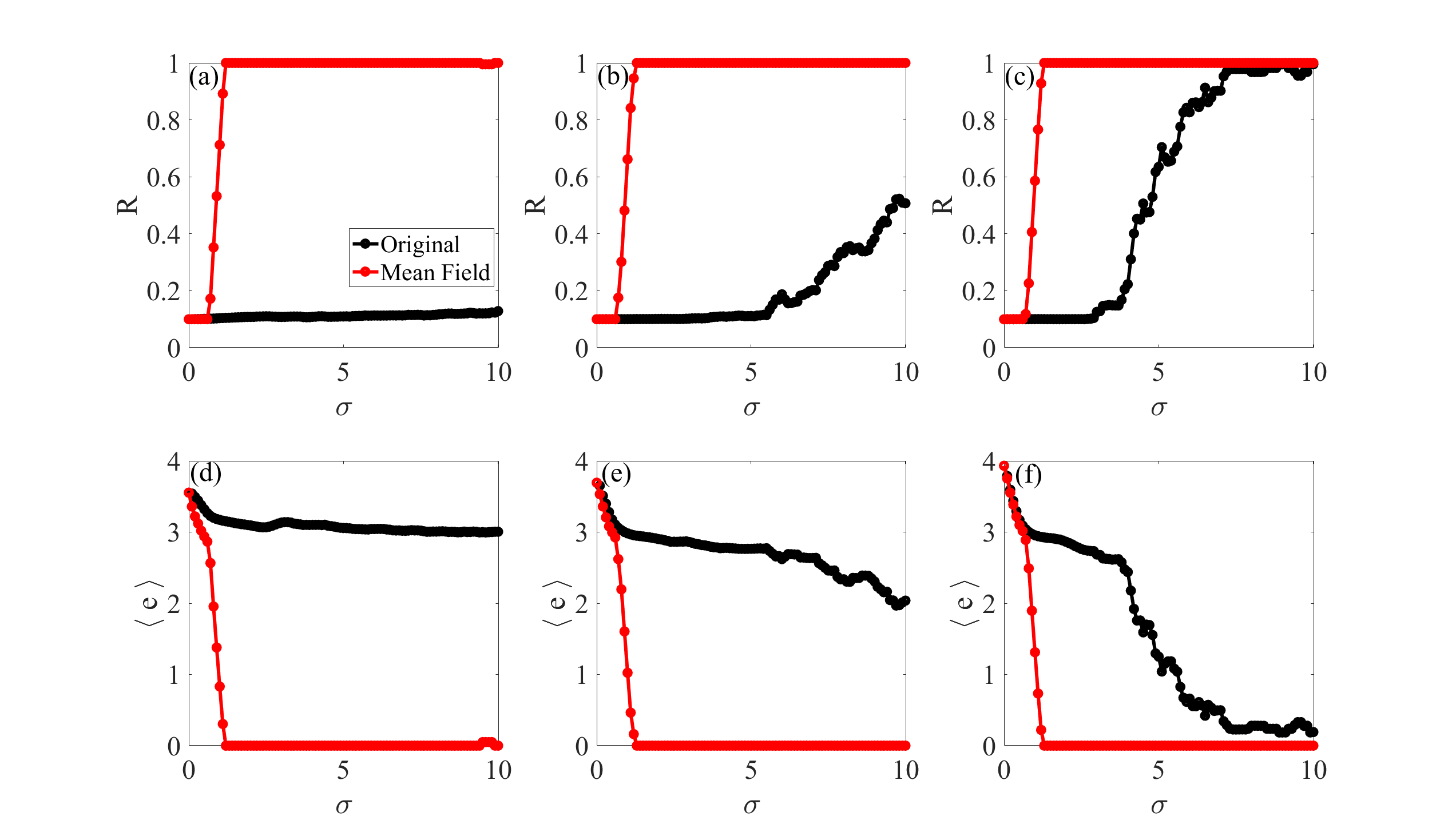}
\end{tabular}
\caption{Synchronization transitions in a heterogeneous network with fitness parameter $\delta = 35\times 10^{-3}$ and identical oscillators as a function of the coupling strength~$\sigma$. The top row (a--c) shows the average order parameter, while the bottom row (d--f) shows the average synchronization error. Columns from left to right correspond to $d = 0.8$, $1.0$, and $1.2$, respectively.}
\label{fig::ER_fullper}
\end{figure}

We now turn to the regime of strong structural heterogeneity, obtained by setting the fitness parameter to a small value, $\delta = 35\times 10^{-3}$. In this configuration, the network could becomes fragmented and exhibits pronounced differences in node connectivity. In this vein, Fig.~\ref{fig::ER_fullper} illustrates the impact of this stronger topological heterogeneity on the synchronization transitions, for the same values of $d$ as those considered in Fig.~\ref{fig::MER_full} and Fig.~\ref{fig::ER_full}. A quick comparison between Fig.~\ref{fig::MER_full}, obtained for $\delta = 10^{5}$ and corresponding to an almost fully connected network, Fig.~\ref{fig::ER_full}, obtained for $\delta = 1$ and yielding a random-graph topology, and Fig.~\ref{fig::ER_fullper}, obtained for $\delta = 35\times 10^{-3}$ and possibly leading to a fragmented network, highlights the crucial role of topological heterogeneity in shaping collective synchronization.
An important conclusion from this analysis is that the homogeneous mean-field (HMF) approximation is reliable only when the underlying network is sufficiently dense and connected, while it progressively loses accuracy as the topology becomes more heterogeneous and fragmented. In particular, the HMF model reproduces well the dynamics of the original system in the (almost) fully connected regime, as shown in Fig.~\ref{fig::MER_full}. However, as soon as the network becomes heterogeneous, the HMF description starts to underestimate the system dynamics (see Fig.~\ref{fig::ER_full}), and a clear mismatch emerges between the mean-field predictions and the full network simulations. This discrepancy becomes even more pronounced in Fig.~\ref{fig::ER_fullper}, where the agreement between both approaches is almost completely lost. In this strongly fragmented regime, the HMF model predicts synchronization behaviors that are not observed in the original system. This failure can be naturally explained by the fragmentation of the network: nodes belonging to disconnected components cannot coordinate their dynamics, and therefore global synchronization cannot be established across the whole network.

\begin{figure}[htp!]
\centering
\begin{tabular}{c}
\includegraphics[width=0.425\textwidth]{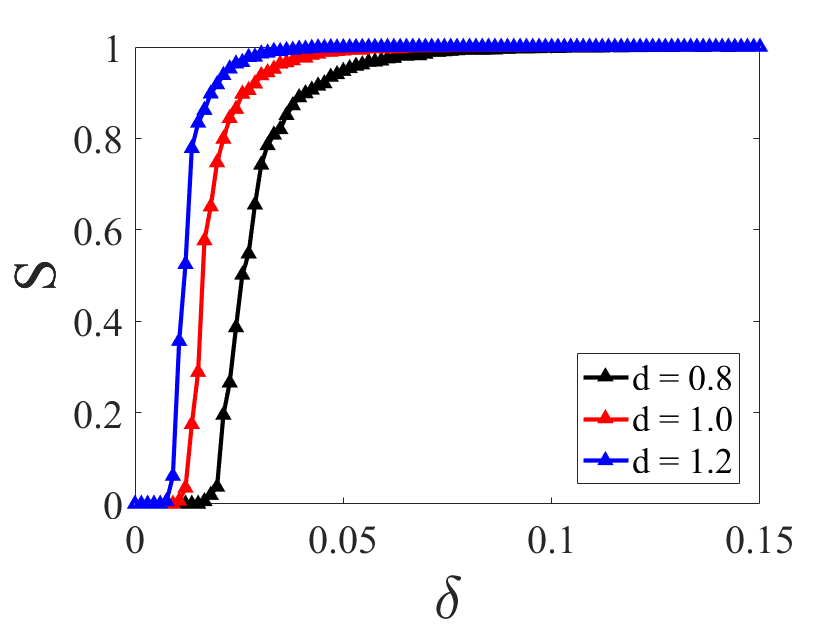}
\end{tabular}
\caption{Formation of the giant component as a function of $\delta$ for three values of  $d = 0.8$, $1.0$, and $1.2$ respectively.}
\label{fig::Giantcomp}
\end{figure}

\begin{figure}[htp!]
\centering
\begin{tabular}{c}
\hspace{-0.550 cm}
\includegraphics[width=0.550\textwidth]{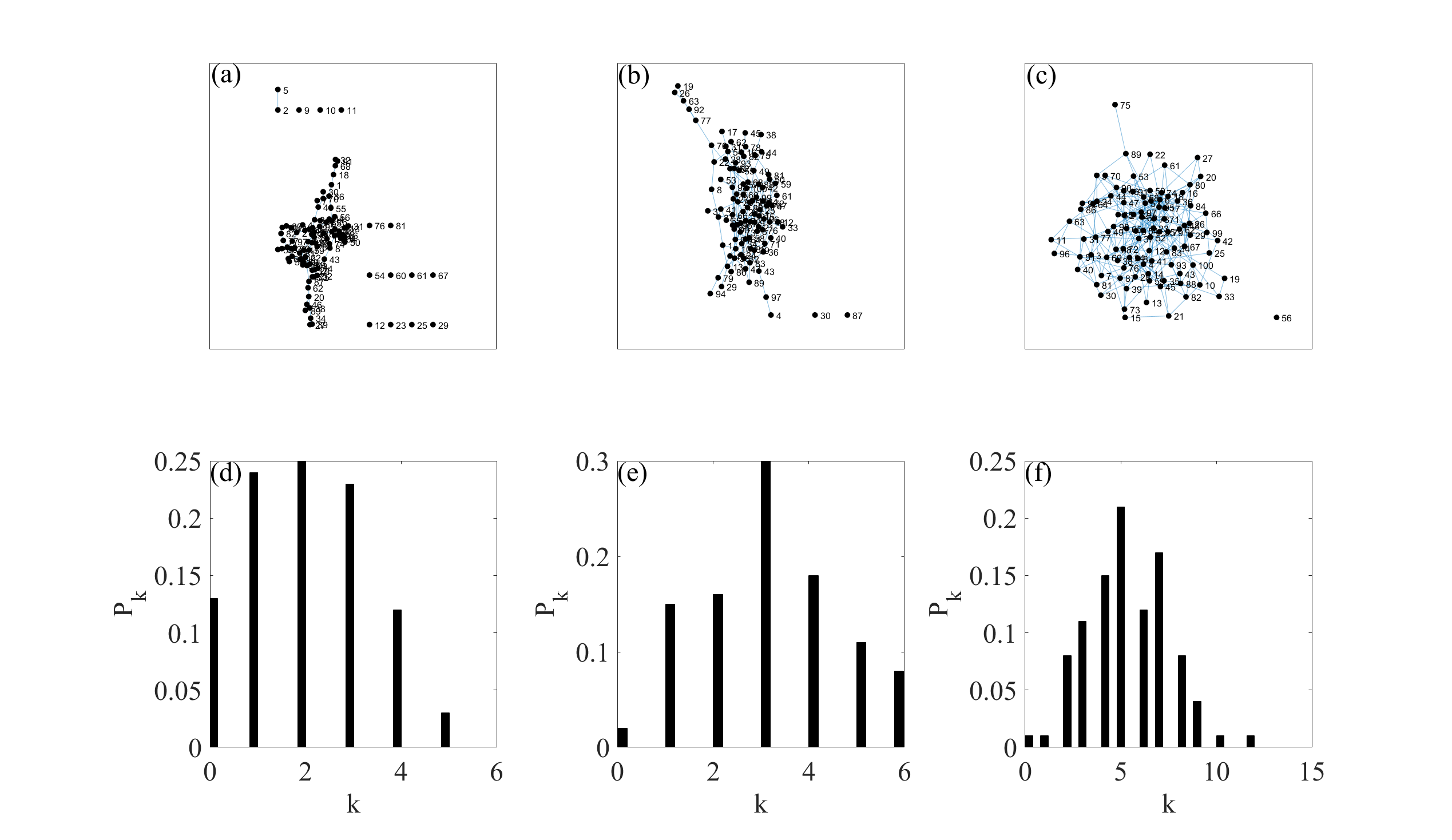}
\end{tabular}
\caption{Variation of the network topology with the parameter $d$. The top row shows the graph representations, and the bottom row the corresponding degree distributions for (a,d) $d = 0.8$, (b,e) $d = 1.0$, and (c,f) $d = 1.2$, with a fitness parameter $\delta = 35\times 10^{-3}$. Higher values of $d$ could correspond to economies with greater productive potential, resulting in denser and more interconnected network structures.}
\label{fig::ER_graph}
\end{figure}

Let us now investigate the regime where the parameter $\delta$ is sufficiently small such that the probability of creating a link between two nodes $j$ and $k$ remains of order $1/N$. In this limit, one can estimate the corresponding critical threshold $\delta_c$ as: 
\begin{equation}
    \delta_c \approx \frac{1}{d^2 N}
\end{equation}
Below this critical threshold, the network remains fragmented into small components, possibly including isolated nodes, and the economy (system) behaves as a collection of disconnected economies with limited trade or financial interactions. In contrast, for $\delta>\delta_c$, a giant connected component emerges~\cite{molloy1995critical} and progressively grows until the network becomes connected. This can be interpreted as the formation of a globally integrated economic system, where a sufficient level of mutual connectivity and trust enables large-scale coordination among economies. This behavior is clearly illustrated by the evolution of the giant component size $S$ in Fig.~\ref{fig::Giantcomp}, which also highlights the dependence of the critical threshold $\delta_c$ on the parameter $d$.  
This transition corresponds to a typical percolation process: as $\delta$ increases, initially disconnected clusters merge into larger connected structures, ultimately forming a macroscopic component spanning a finite fraction of the network, as illustrated in Fig.~\ref{fig::ER_graph}.

To better understand the results in Fig.~\ref{fig::ER_fullper}, where a strong discrepancy is observed between the HMF approach and the original model, we report in Fig.~\ref{fig::ER_graph} three typical network realizations obtained for $\delta = 35 \times 10^{-3}$ and for three values of $d=\{0.8,\,1.0,\,1.2\}$. As discussed above, the percolation threshold $\delta_c$ depends on $d$, which directly affects the global connectivity of the network and, consequently, the onset of synchronization.
For the first two configurations, corresponding to $d=0.8$ and $d=1.0$ respectively, Fig.~\ref{fig::ER_graph} clearly reveals the presence of isolated nodes and disconnected components. In such a fragmented scenario, local interactions cannot propagate through the entire network, preventing the establishment of global phase coherence. From an economic perspective, these cases can be interpreted as poorly integrated systems, where weak trade or trust connections hinder large-scale coordination of business cycles, even under strong coupling.
In contrast, for $d=1.2$, the network becomes almost fully connected, with a giant component size approaching unity, i.e., $S\to 1$. This explains the emergence of a large order parameter and the corresponding synchronized regime observed in the original model for large $\sigma$.  Indeed, the mean-field approximation fails to capture the impact of structural heterogeneity and predicts an almost immediate transition to phase and full synchronization at very small values of $\sigma$. This systematic overestimation highlights a fundamental limitation of the HMF framework in heterogeneous networks, where the assumption of homogeneous mixing is no longer justified.
Within the economic analogy, this approximation implicitly assumes nearly perfect global integration and uniformly distributed trust, thereby overpredicting collective stability.

\subsection{Heterogeneous economic network with distributed potential GDPs} \label{sec::subsec3C}

\begin{figure}[htp!]
\centering
\begin{tabular}{c}
\hspace{-1.0 cm}
\includegraphics[width=0.550\textwidth]{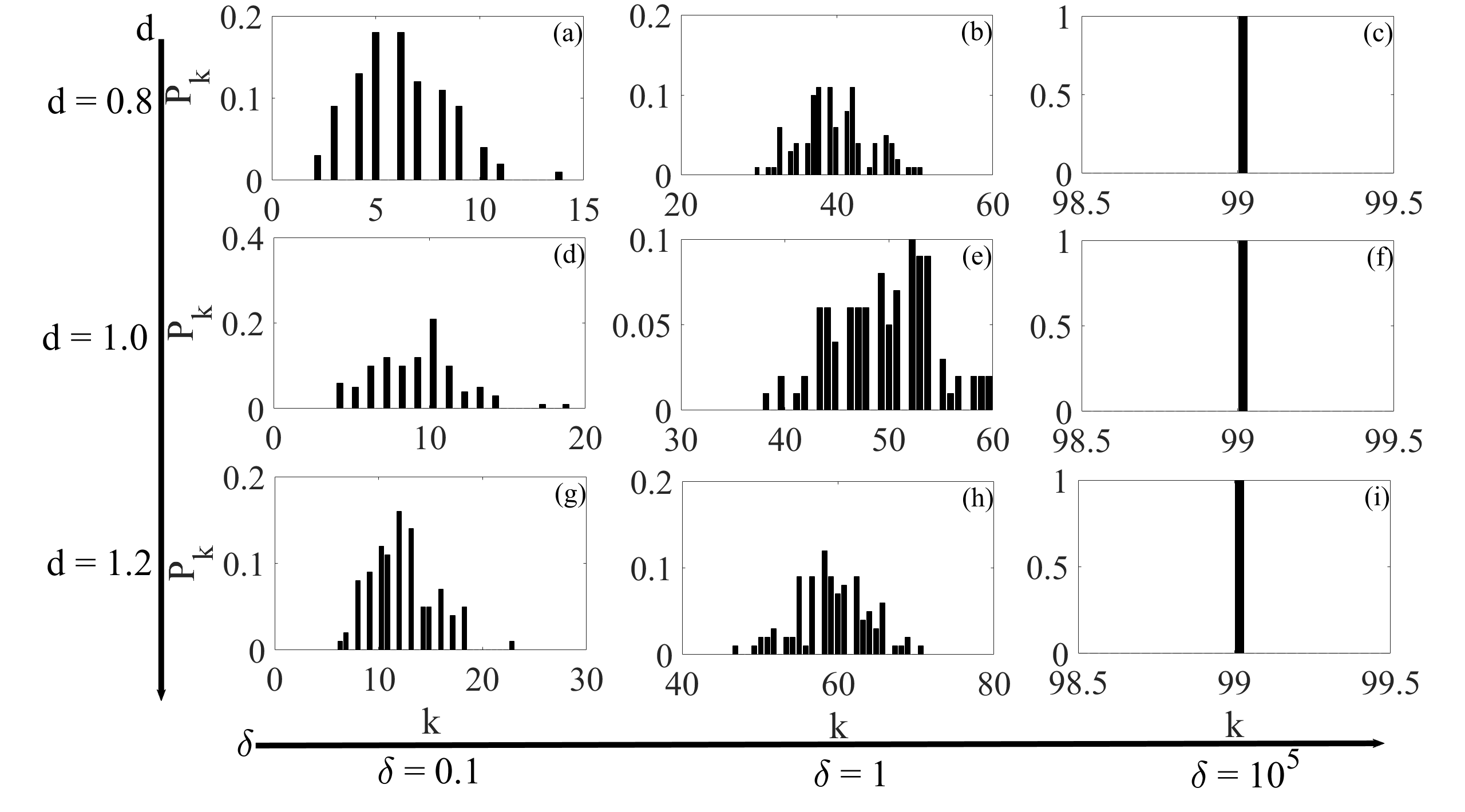}
\end{tabular}
\caption{Degree distribution of the network for different potential GDP values, $\bar d \in \{0.8,\,1.0,\,1.2\}$, and fitness parameters, $\delta \in \{10^{-1},\,1.0,\,10^{5}\}$, showing how these factors shape network connectivity for $\chi = 0.1$.
}
\label{fig::disthetero}
\end{figure}

We now revisit the network of $N$ macroeconomic oscillators governed by Eq.~\ref{eq::Cam_het}. In this formulation, each unit is intrinsically heterogeneous, being characterized by its own parameter value $d_j$ ($j = 1,2,\dots,N$). These parameters are independently drawn from a log-normal distribution such that
$
\ln d_j \sim \mathcal{N}(\bar d,\chi^2),
$
with a mean $\bar d$ and a standard deviation $\chi$ as defined in Sec.~\ref{sec::subsec31}. This distribution of $d$ across the nodes determines the heterogeneous structure of the network.  Such heterogeneity manifests itself at two distinct levels: (i) structural, since differences in $d_j$ directly affect the connection probability between nodes and hence the overall network topology (See Fig.~\ref{fig::disthetero}; the comments are similar to those for Fig.~\ref{fig::disthet}.); and (ii) dynamical behavior, as each system follows an individual trajectory depending on its own parameter $d_j$. 

\begin{figure}[htp!]
\centering
\begin{tabular}{c}
\includegraphics[width=0.45\textwidth]{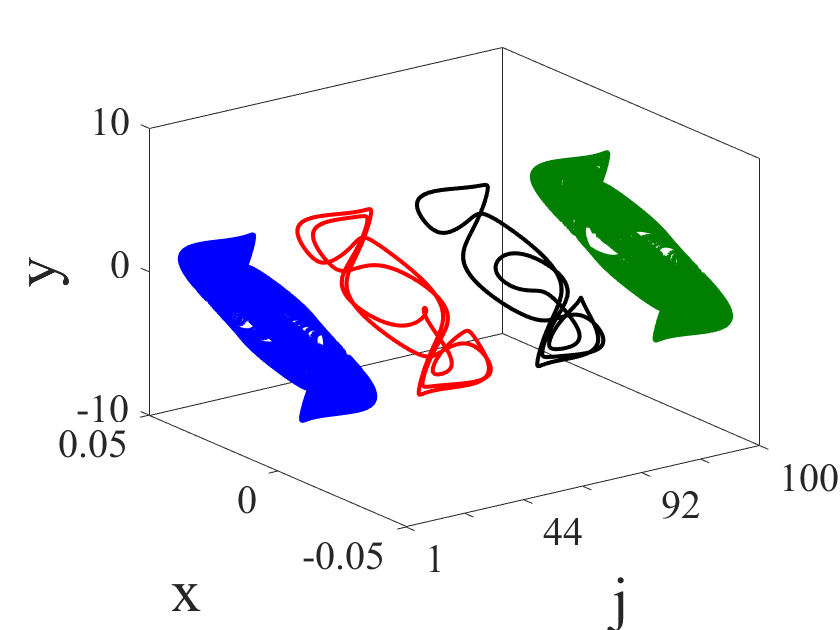}
\end{tabular}
\caption{Phase portraits of four selected oscillators from the network, where the parameters $d_j$ are drawn from a log-normal distribution $\ln d \sim \mathcal{N}(\bar{d}, \chi^2)$ with $\bar{d}=1$ and $\chi=0.025$. The coupling is set to zero ($\sigma=0$), so that each oscillator evolves independently according to its intrinsic dynamics, representing isolated economies with no mutual interactions.}
\label{fig::Port}
\end{figure}

As an illustration of point~(ii), Fig.~\ref{fig::Port} displays the intrinsic dynamics of four representative oscillators in the network, labeled $j \in \{1, 44, 92, 100\}$, in the absence of coupling, i.e., for $\sigma = 0$ and as a consequence, $P_k = 0$.
Each node $j$ evolves independently according to its value of the parameter $d$, drawn from the distribution defined above. The resulting phase portraits exhibit diverse dynamical regimes—ranging from multi-periodic to chaotic—reflecting the coexistence of different attractors across isolated systems (or economies), even under a very weak dispersion of the parameter $d$, with the variability fixed at $\chi = 0.025$. In this uncoupled state, each economy follows its own endogenous cycle of expansion and contraction, with no coordination or mutual influence.

Consequently, when interactions between systems are introduced (i.e., $\sigma \neq 0$ and then, $P_k \ne 0$), the dispersion parameter $\chi$ affects the degree of heterogeneity in the network: larger values of $\chi$ induce stronger variability across units, giving rise to configurations in which an increasing number of subsystems exhibit multi-periodic dynamics, while others display chaotic behavior. This dual heterogeneity introduces both topological and dynamical diversity, which strongly influence the emergence and stability of the collective behavior and, especially, the synchronization. From economic point of view, it  reflects a global system of economies with heterogeneous productive capacities and growth potentials, where structural disparities in connectivity and intrinsic dynamics shape the overall pattern of macroeconomic coordination.

\begin{figure*}[htp!]
\centering
\begin{tabular}{c}
\hspace{0.0 cm}
\includegraphics[width=0.950\textwidth]{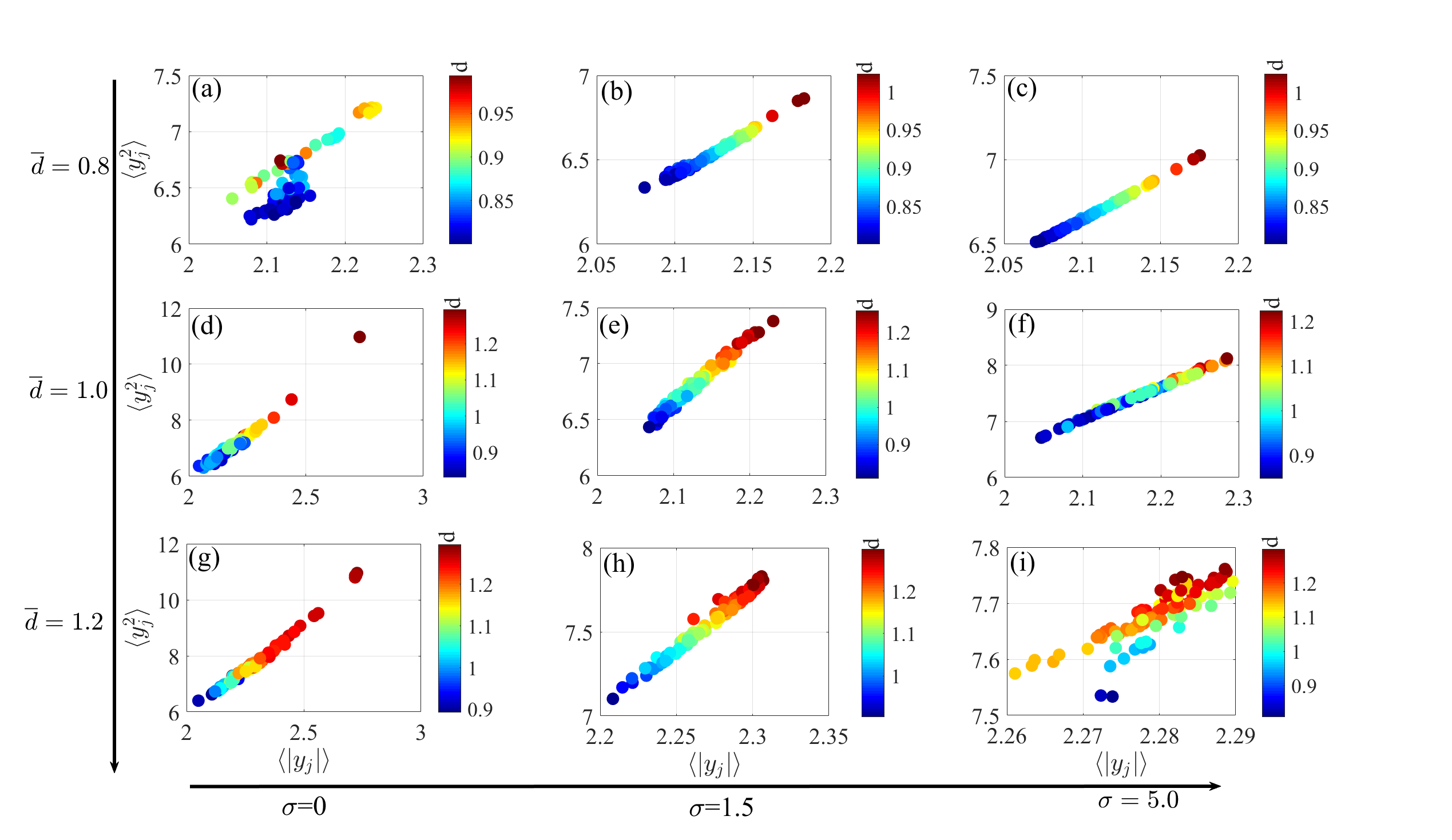}
\end{tabular}
\caption{Relationship between the time-averaged squared GDP, $\langle y_j^2 \rangle$, and the time-averaged absolute GDP, $\langle |y_j| \rangle$, for all nodes in a highly heterogeneous network with $\delta = 1$ (see panels~\ref{fig::disthetero}(b,e,h) in Fig.~\ref{fig::disthetero}). The first row (a–c) corresponds to $\bar d = 0.8$, the second row (d–f) to $\bar d = 1.0$, and the third row (g–i) to $\bar d = 1.2$. From left to right, the columns represent: (a,d,g) an uncoupled network ($\sigma = 0$), (b,e,h) $\sigma = 1.5$, and (c,f,i) $\sigma = 5$.  The parameter $d$ is drawn from a log-normal distribution over the interval $[0.8, 1.3]$, with $\ln d \sim \mathcal{N}(\bar d, \chi^2)$ and $\chi = 0.1$.
}
\label{fig::intsync}
\end{figure*}

In order to verify and further support the choice of the parameter $d$ to evaluate the connection probability between systems, Fig.~\ref{fig::intsync} illustrates the relationship between $\langle y_j^2 \rangle$, the time average of $y_j^2$ over a sufficiently long time window $T$, and $\langle |y_j| \rangle$, the time average of $|y_j|$ over the same interval, in a fully heterogeneous network that is heterogeneous both dynamically and topologically. The dynamical heterogeneity arises from the log-normal distribution of $d_j$, while the topological heterogeneity corresponds to $\delta = 1$, which generates a broad degree distribution as shown in Fig.~\ref{fig::disthetero}(b,e,h). The analysis is performed for three values of $\bar d \in \{0.8,\,1.0,\,1.2\}$. Despite the pronounced heterogeneity of the network, the results displayed in Fig.~\ref{fig::intsync} reveal a strong and systematic correlation between $\langle y_j^2 \rangle$ and $\langle |y_j| \rangle$ in all cases, consistent with the behavior observed in Fig.~\ref{fig::Cor_dy}. 
This result further corroborates and reinforces the relevance of the parameter $d$ as an appropriate quantity for defining the connection probability.

\subsubsection{Homogeneous mean-field approximation in the fully connected heterogeneous economic network} \label{sec::subsubsec3C1}

\begin{figure}[htp!]
\centering
\begin{tabular}{c}
\hspace{-0.80 cm}
\includegraphics[width=0.57\textwidth]{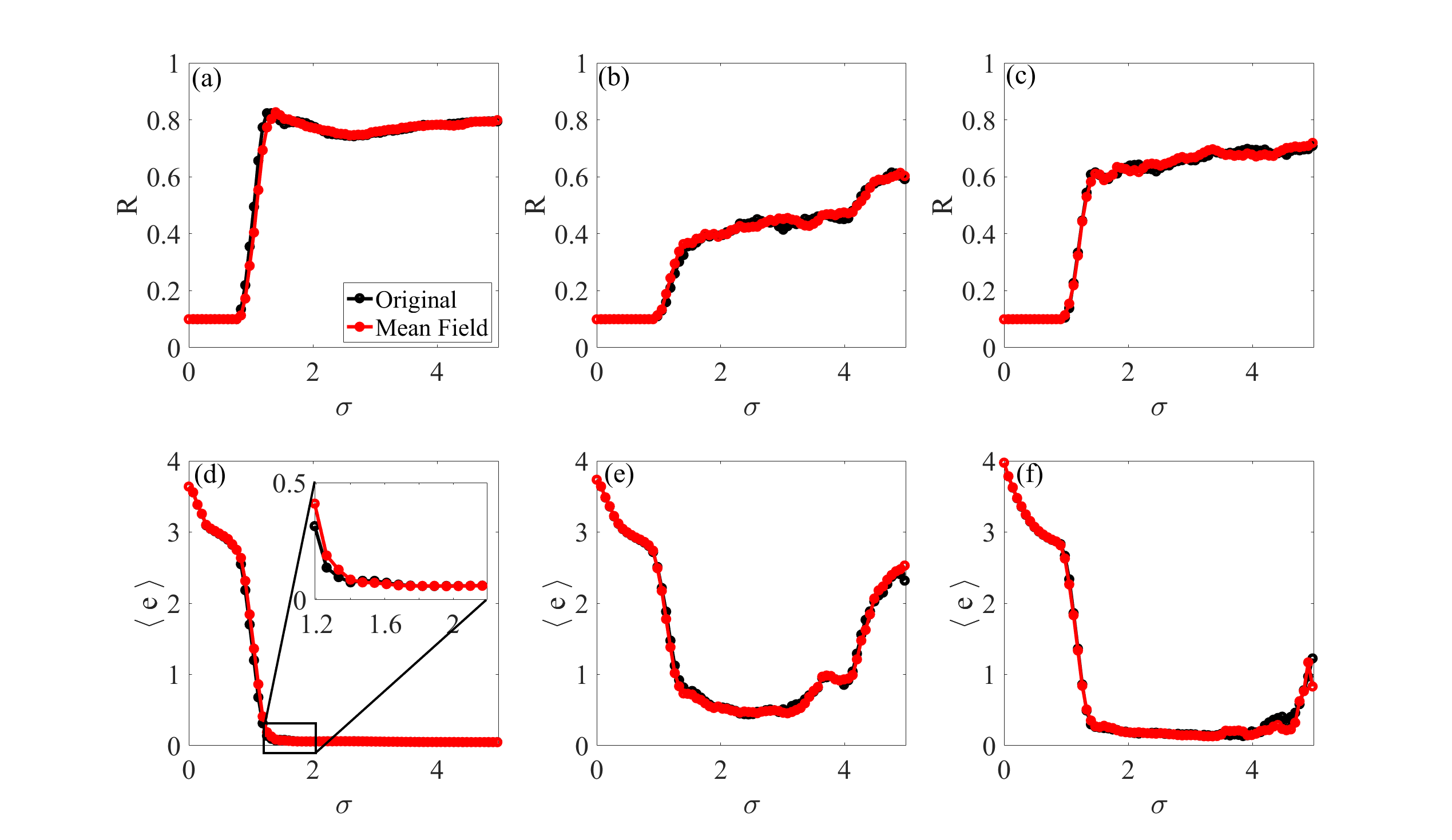}
\end{tabular}
\caption{Synchronization transitions in a heterogeneous network in a fully connected topology ($\delta = 10^5$) of nonidentical oscillators as a function of the coupling strength~$\sigma$. Top row (a--c): average order parameter. Bottom row (d--f): average synchronization error. Columns correspond to $\bar{d} = 0.8$, $1.0$, and $1.2$, respectively with variability $\chi=0.1$.}
\label{fig::Hetero_graph02}
\end{figure}

We now examine the case of a fully connected network ($\delta = 10^5$, as in Fig.~\ref{fig::MER_full}), in which all economies are uniformly connected and heterogeneity arises solely from their intrinsic dynamics, without any structural asymmetry in connectivity. The main goal here is to assess the validity of the HMF framework developed earlier in a more specific setting, where the network remains fully connected but exhibits heterogeneity stemming from the internal dynamics of each system. For this purpose, we analyze the model formulated under the HMF framework in Eq.~\ref{eq::homomff}, where the parameter $d$ for each node $j$ is sampled from the log-normal distribution introduced earlier, with values constrained to the range $d_j \in [0.8, 1.3]$. A numerical analysis performed under the same initial conditions and parameter values as in Sec.~\ref{sec::subsubsec3B1}, enables us to reassess the synchronization transition (see Fig.~\ref{fig::Hetero_graph02}) for this network of heterogeneous macroeconomic systems.

The synchronization properties of the network are summarized in Fig.~\ref{fig::Hetero_graph02}. Panels ~\ref{fig::Hetero_graph02}(a–c) show the average order parameter $R$, while panels~\ref{fig::Hetero_graph02}(d–f) display the average synchronization error $\langle e \rangle$. From left to right, the columns correspond to $\bar{d} = 0.8$, $1.0$, and $1.2$, respectively, with a variability of $\chi = 0.1$. Despite the intrinsic heterogeneity among systems, both the original model (black curves) and the mean-field approximation (red curves) exhibit almost identical behavior. As the coupling strength $\sigma$ increases, the system undergoes a sharp transition from an incoherent regime—characterized by low values of the order parameter ($R \to 0$) and large synchronization errors ($\langle e \rangle$ high)—to a partially synchronized state ($0 < R < 1$). The excellent agreement between the two approaches indicates that, in a fully connected network, structural homogeneity (i.e., $P_k=N-1$ for all nodes) dominates the effects of dynamical heterogeneity. This agreement confirms that when all nodes—or economies—interact equally with one another, local differences in intrinsic dynamics are effectively averaged out by the uniform coupling. This behavior remains essentially unchanged even when the variability of $d$ is increased (see Fig.~\ref{fig::Hetero_graph02_app} in Appendix~\ref{sec::app1}), confirming that structural homogeneity continues to dominate over dynamical diversity. It should be emphasized that in the heterogeneous case, where diversity arises both from the intrinsic dynamics of the systems and from the network topology, the mean-field approximation fails to accurately capture the collective behavior. As illustrated in Fig.~\ref{fig::Hetero_graph01_app} in Appendix~\ref{sec::app2}, this approach breaks down when connectivity is unevenly distributed among nodes. 
 
\begin{figure}[htp!]
\centering
\begin{tabular}{c}
\hspace{-1.0 cm}
\includegraphics[width=0.55\textwidth]{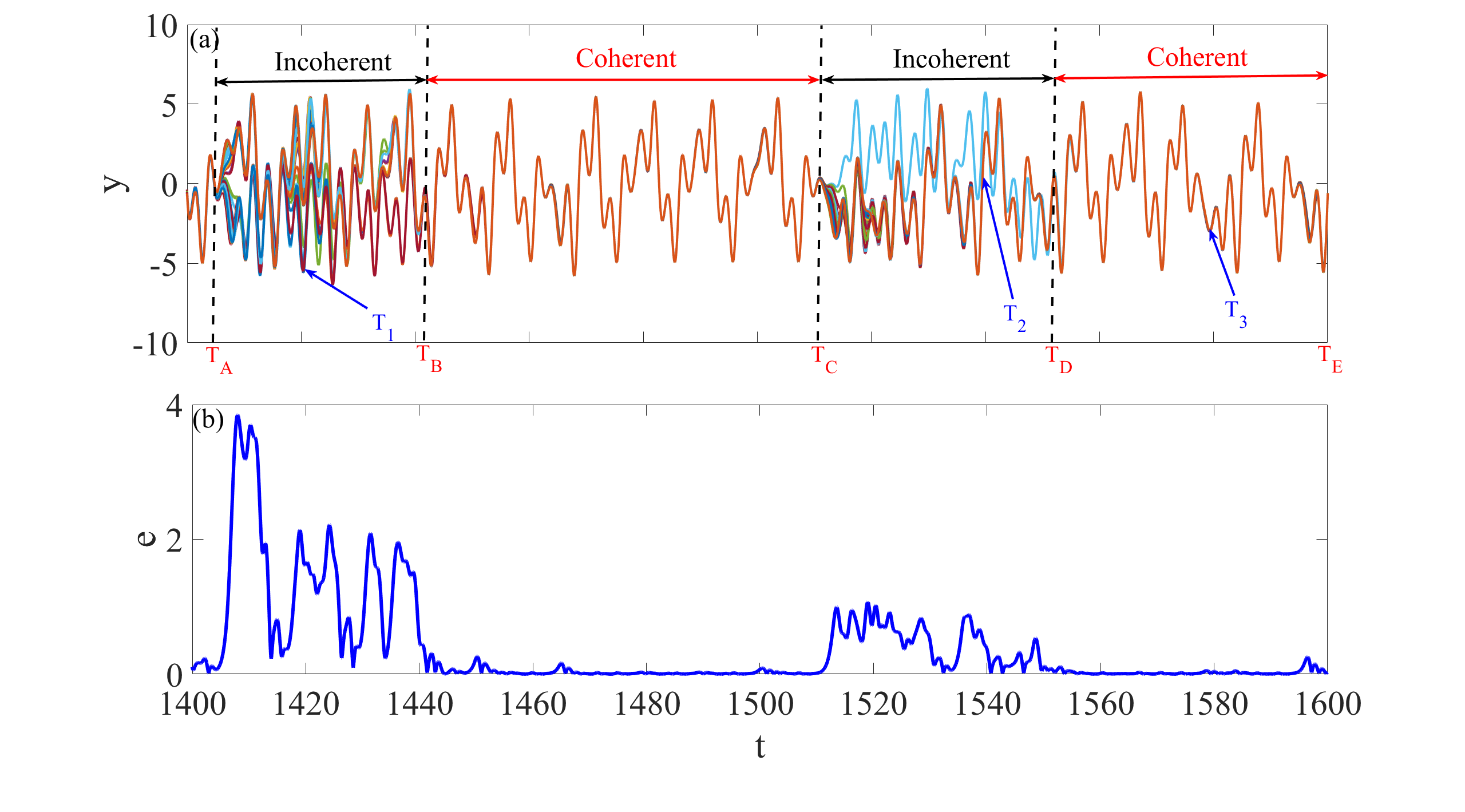}
\end{tabular}
\caption{(a) Time series of the GDP variable $y$, illustrating intermittent synchronization; 
(b) synchronization error; and 
The parameters are: $\bar{d} = 1$, with $\chi = 0.1$, $\sigma = 2.5$, and $\delta = 10^{5}$.}
\label{fig::Time_y_E}
\end{figure}

In contrast to the homogeneous network case, where all entities were identical, an important observation here is that the order parameter does not reach its maximal value $R = 1$, which would correspond to perfect phase synchronization. To explore this phenomenon further, Fig.~\ref{fig::Time_y_E} displays in panel~\ref{fig::Time_y_E}(a) the temporal evolution of the GDP variable $y(t)$ for each node in the network, along with the corresponding synchronization error in panel~\ref{fig::Time_y_E}(b).
The analysis of these quantities reveals that the network does not exhibit a state of permanent synchronization or total desynchronization. Instead, the system displays \textit{intermittent synchronization}, where the trajectories alternate between coherent and incoherent phases~\cite{boccaletti2002synchronization}.

Let us focus here on the nature of the transitions occurring during the alternating synchronized and desynchronized phases illustrated in Fig.~\ref{fig::Time_y_E}(a). At each onset of desynchronization (see the intervals $[T_A, T_B]$ and $[T_C, T_D]$), such as at points $T_A$ and $T_C$, one observes that the loss of coherence arises suddenly and spontaneously. Prior to these instants, the systems remain tightly synchronized, evolving collectively along a common trajectory. The abrupt divergence appearing at $T_A$ and $T_C$ therefore resembles a spontaneous instability or an external shock that instantaneously disrupts the synchronization. In contrast to these abrupt desynchronization events, the transition toward synchronization develops gradually, as observed in the time intervals $[T_B, T_C]$ and $[T_D, T_E]$ mark as coherents. During these periods, coherence emerges progressively through the temporary formation of synchronized clusters which merge until the entire network reaches a quasi-synchronous state. Such asymmetry between the slow formation of synchronization and its abrupt collapse is a characteristic feature of \textit{on--off intermittency}~\cite{platt1993off,koronovskii2024intermittent}. This behavior reflects the metastable~\cite{caprioglio2024emergence} nature of the synchronized manifold, which corresponds to a state that appears stable over long time intervals but remains only weakly attracting. In other words, the system can remain in the synchronized state for extended periods, giving the impression of stability, yet small perturbations can easily destabilize it, leading to sudden desynchronization. This weakly stable character underlies the observed intermittent transitions between synchronized and desynchronized phases. As a result, desynchronization can be triggered explosively by infinitesimal perturbations, whereas the recovery of synchronization requires a much slower process of phase adjustment and gradual re-entrainment. From economical point of view, it highlights how global coordination among economies is often destroyed by sudden crises or shocks, but rebuilt progressively through phases of regional alignment, trade reinforcement, and financial interdependence.

\begin{figure}[htp!]
\centering
\begin{tabular}{c}
\hspace{-1.0 cm}
\includegraphics[width=0.57\textwidth]{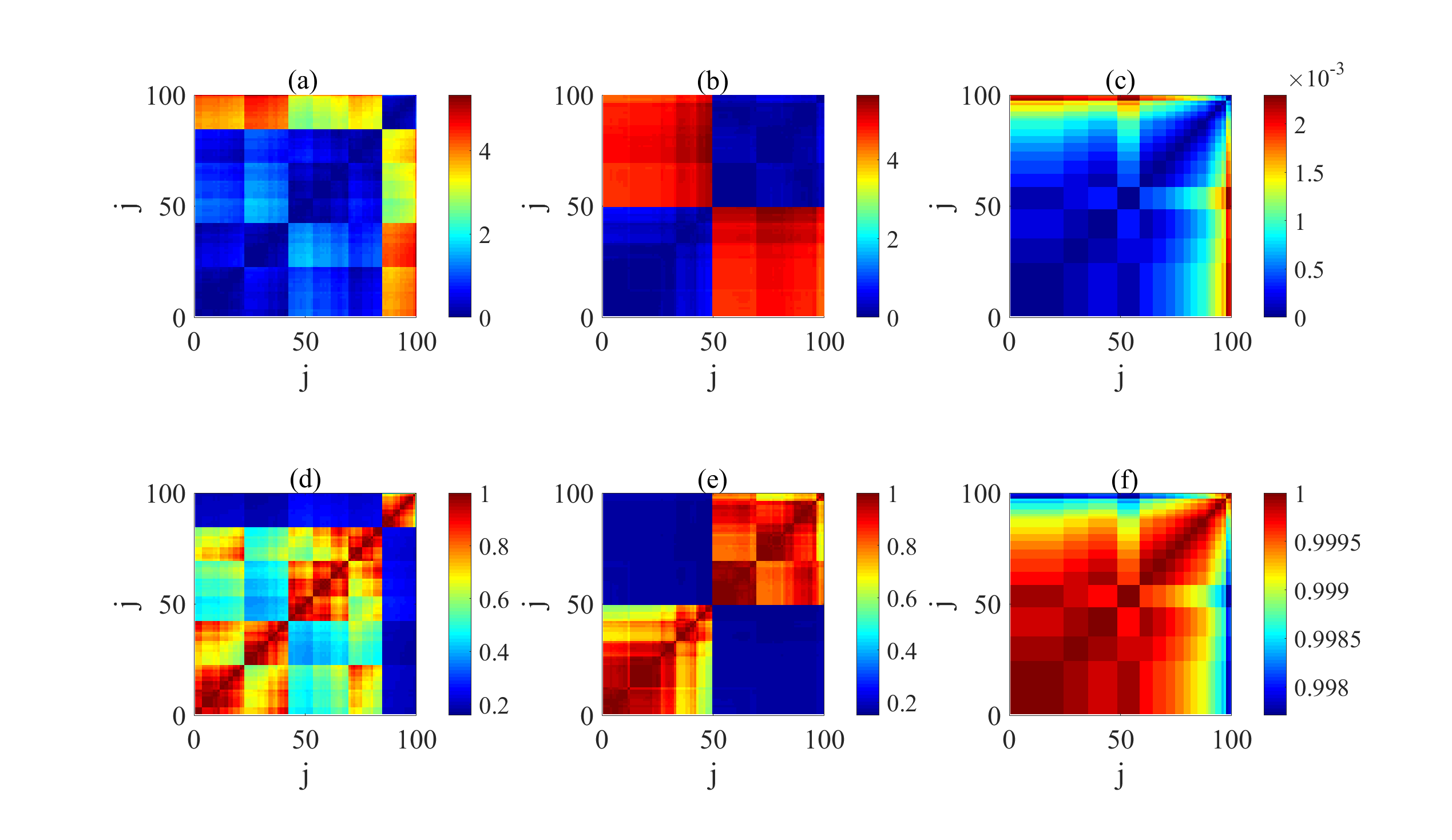}
\end{tabular}
\caption{Snapshots showing (a--c) the pairwise distance $D_{ij}$ and (d--f) the normalized similarity $J_{ij}$ between systems in the network. From left to right, the panels correspond to (a,d) $t = 1421$, (b,e) $t = 1540$, and (c,f) $t = 1580$. The parameters are the same as those used in Fig.~\ref{fig::Time_y_E}. The sequence illustrates the progressive formation and merging of synchronized clusters leading to full network synchronization, corresponding to the transition from fragmented to globally coherent economic dynamics.}
\label{fig::RP}
\end{figure}

To further analyze how the network transitions between desynchronization, cluster formation, and full phase synchronization in Fig.~\ref{fig::Time_y_E}, we examine the time series of the pairwise distances $D_{ij}$ and similarities $J_{ij}$. These quantities, computed from the GDP variable $y$ according to Eq.~\ref{eq::DS}, are shown in Fig.~\ref{fig::RP}.
\begin{equation}
    D_{ij}=|\mathbf{X}_i - \mathbf{X}_j|, \quad  \text{and} \quad J_{ij} = \frac{1}{1+D_{ij}}
    \label{eq::DS}
\end{equation}
As the system evolves from desynchronization to full synchronization, the pairwise distance between oscillators decreases, and the similarity between them increases, reflecting the convergence of their dynamics. 

The point $T_1 = 1421$ on the time series of the variable $y$, which occurs after the sudden desynchronization at $T_A$, marks the emergence of synchronized clusters. Although the system does not exhibit perfect synchronization, the emergence of clusters is already noticeable, as shown in panel~\ref{fig::RP}(a,d). At this stage, economies begin to collectively reorganize after the shock occur at $T_A$, with synchronized subgroups starting to form. Initially, a large number of clusters emerge, which gradually reduce over time, passing through a state with two clusters as represented in panel~\ref{fig::RP}(b,e) at $T_2 = 1540$) where it is seasy to identify two group, before reaching the final phase of full synchronization as at $T_3 = 1580$, where only one cluster is observed, as shown in panel~\ref{fig::RP}(c,f). This final state reflects a fully coherent network, where all economies have synchronized their dynamics, as indicated by the minimal distance and maximum similarity between them.

This intermittent behavior explains why the order parameter shown in Fig.~\ref{fig::Hetero_graph02} never reaches unity: synchronization is only temporary, as coherent phases are repeatedly interrupted by bursts of desynchronization.  The time series also shows that oscillators initially exhibiting multi-periodic dynamics can, during certain intervals, adopt the chaotic behavior of others, illustrating how less volatile economies may temporarily follow the unstable trajectories of more turbulent ones.

\begin{figure}[htp!]
\centering
\begin{tabular}{c}
\includegraphics[width=0.35\textwidth]{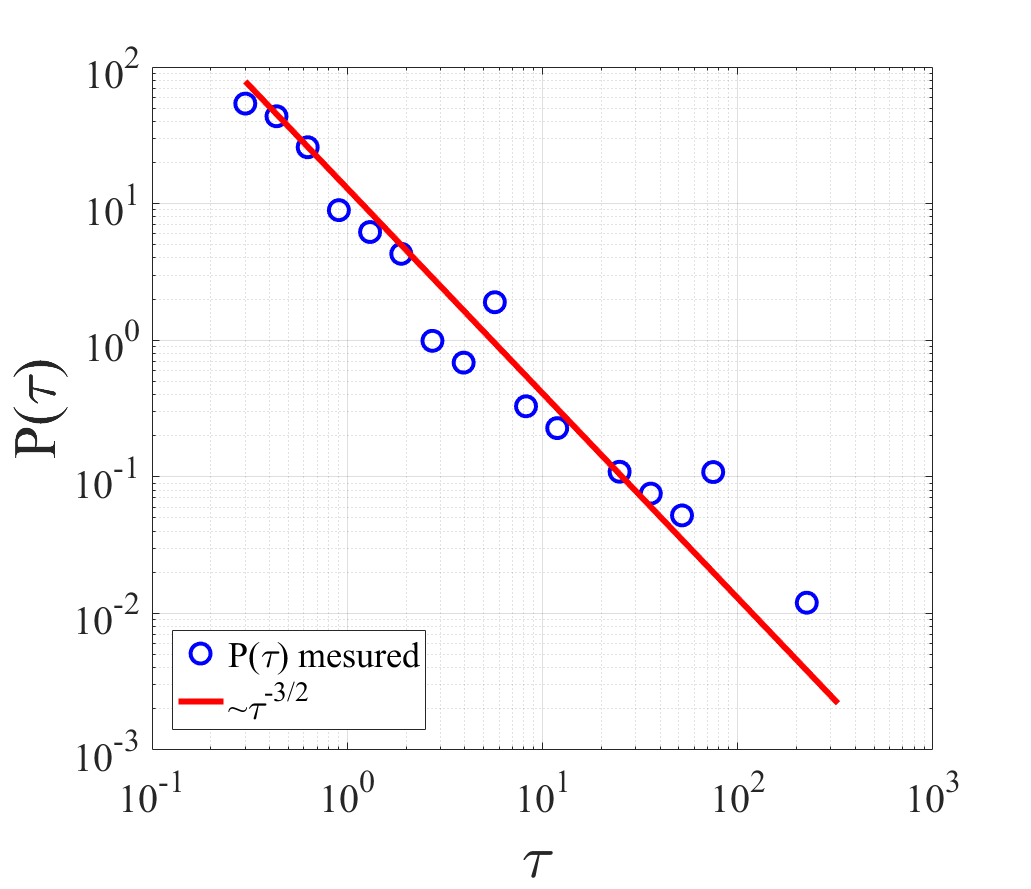}
\end{tabular}
\caption{Probability distribution $P(\tau)$ of the laminar phase durations $\tau$, 
measured during the intermittent synchronization regime. The parameters are $\bar{d} = 1$, $\chi = 0.1$, $\sigma = 2.5$, and $\delta = 10^{5}$.}
\label{fig::TL}
\end{figure}

Let us now quantitatively characterize the intermittency observed in the synchronization process presented in Fig.~\ref{fig::Time_y_E}, which is done by analyzing the distribution of laminar durations~\cite{koronovskii2024intermittent} for very long simulation time, defined as the time intervals during which the system remains in a quasi-synchronized state. These durations are numerically computed for each oscillator (or economy) in the network, and the resulting distribution, $P(\tau)$, is shown in Fig.~\ref{fig::TL}. Theoretically, this distribution follows a power-law decay, $P(\tau) \sim \tau^{-3/2}$, typical of on--off intermittency~\cite{platt1993off}.
The exponent $3/2$ of this power-law provides valuable insight into the nature of synchronization in the system. A larger exponent indicates that shorter laminar phases are more frequent, meaning synchronization is more transient and the system oscillates between coherent and incoherent states more rapidly. In the sens interconnected economies, this implies that global economic alignments (such as synchronized booms or recessions) are short-lived, with economies quickly diverging due to shocks or policy changes. Conversely, a smaller exponent suggests longer laminar phases, with the system spending more time in synchronized states before switching to desynchronization. From economical perspective, this smaller exponent suggests that periods of global economic coordination last longer before being interrupted by divergence.

\subsubsection{Heterogeneous mean-field approximation $\delta \to 1$}
Here we focus on the case of a structurally and dynamically heterogeneous network in which the fitness parameter satisfies $\delta > \delta_c$, ensuring that the system remains connected but exhibits nontrivial degree variability. In this regime, the connection probability between two nodes $j$ and $k$ can be approximated as $P_{jk} \approx \delta d_j d_k$, following the standard assumption of fitness-dependent random networks~\cite{caldarelli2002scale}. Based on this approximation, let us start by deriving the heterogeneous mean-field approximation of the original network model given by Eq.~\ref{eq::Cam_het}. We assume that the microscopic connectivity of each node can be replaced by an effective coupling term that depends on its expected degree. In the original model in Eq.~\ref{eq::Cam_het}, the diffusive coupling term $\frac{\sigma}{k_j}\sum_{k=1}^N A_{jk}(z_k - z_j)$ explicitly depends on the adjacency matrix $A_{jk}$ and the degree $k_j$ of node $j$. Under the Heterogeneous mean-field approximation, local connectivity patterns are replaced by their statistical averages over nodes with similar properties~\cite{boguna2003absence}. In this model, the sum $\sum_{k=1}^N A_{jk} z_k$ is approximated by the expected value :
\begin{equation}
    \sum_{k=1}^N A_{jk} z_k \sim \frac{k_j}{\langle k \rangle} \sum_{k=1}^N P_k z_k
\end{equation}
where $P_k$ represents the probability of connection to a node with potential GDP $d_k$. Remember that, the network is constructed from a  model where the connection probability is proportional to the product $d_j d_k$~\cite{caldarelli2002scale}, this average coupling can be expressed as a weighted mean of the neighboring states, as follow: 
\begin{equation*}
    \langle \frac{A_{jk}z_k}{k_j} \rangle \approx \frac{\sum_{k=1}^N d_k z_k}{\sum_{k=1}^N d_k}.
\end{equation*}
Replacing the explicit adjacency term by this weighted mean yields the HMF form of the coupling dynamics shown in Eq.~\ref{eq::heteromf}.
\begin{equation}
    \begin{cases}
        \dot{x}_j = my_j + px_j(d_j - y_j^2), \\
        \dot{y}_j = vy_j + wx_j + cz_j, \\
        \dot{z}_j = sx_j - ry_j + \sigma \left(\frac{\sum_{k=1}^N d_kz_k}{\sum_{k=1}^N d_k} - z_j \right),
    \end{cases}
    \label{eq::heteromf}
\end{equation}

In the present formulation, each node interacts not with specific neighbors but with a global mean field weighted by the economic potential $d_k$ of all other economies. This form preserves the effect of structural heterogeneity—through the distribution of $d_j$ ($\forall j=1,...,N$)—while simplifying the dynamics to a tractable mean-field form suitable for analytical and numerical exploration of synchronization transitions.

\begin{figure}[htp!]
\centering
\begin{tabular}{c}
\hspace{-0.80 cm}
\includegraphics[width=0.55\textwidth]{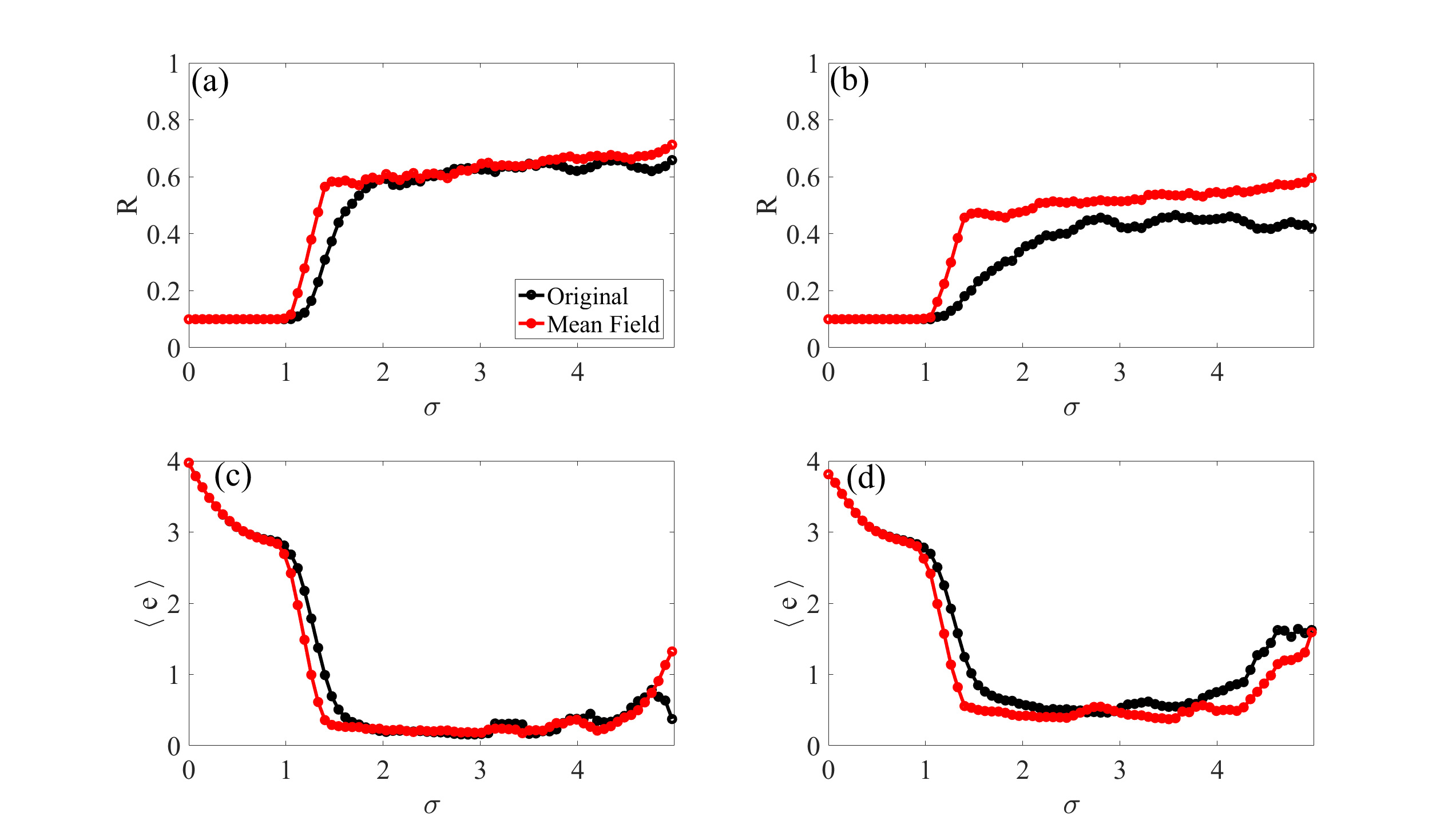}
\end{tabular}
\caption{Synchronization transitions in a heterogeneous network ($\delta = 1$) of nonidentical oscillators as a function of the coupling strength~$\sigma$. The values of $d_j$ are drawn from a log-normal distribution with mean $\bar{d} = 1.2$ and two standard deviations: (a,c) $\chi = 0.1$ and (b,d) $\chi = 0.45$. Top panels (a,b) show the average order parameter, while bottom panels (c,d) display the average synchronization error. The first column corresponds to low variability ($\chi = 0.1$), and the second to higher variability ($\chi = 0.45$).}
\label{fig::Hetero_perco2}
\end{figure}

Fig.~\ref{fig::Hetero_perco2} illustrates the synchronization transitions in a heterogeneous network of nonidentical oscillators, where the structure is fixed ($\delta = 1$) and heterogeneity arises solely from the same log-normal distribution of  $d_j$ with mean $\bar{d} = 1.2$ and two different levels of variability: $\chi = 0.1$ (low heterogeneity) and $\chi = 0.45$ (high heterogeneity). The top row of Fig.~\ref{fig::Hetero_perco2} shows the evolution of the average order parameter $R$ as a function of the coupling strength $\sigma$, while the bottom row presents the corresponding average synchronization error $\langle e \rangle$. Panels~(a,c) correspond to the low-variability case, and panels~(b,d) to the higher-variability regime. \\
As the coupling strength $\sigma$ increases, both models—the original system (black curves) and the heterogeneous mean-field approximation (red curves)—exhibit a clear transition from incoherence to partial synchronization. For both higher and lower variabilities of the $d$ (i.e., $\chi = 0.45$ and $\chi = 0.1$, respectively), corresponding to stronger and weaker structural and dynamical heterogeneity, synchronization remains incomplete: $R$ saturates at lower values and $\langle e \rangle$ does not vanish, indicating persistent incoherence among nodes. The heterogeneous mean-field model reproduces the overall trend of the full dynamics, with nearly identical critical coupling strengths $\sigma$ in both cases, although small deviations appear when heterogeneity becomes significant, reflecting the growing influence of local fluctuations neglected by the mean-field approximation.

\section{Conclusion}\label{sec::con}

This paper has investigated synchronization transitions in a network of coupled chaotic macroeconomic systems. In this framework, the network topology is shaped by a fitness mechanism related to potential GDP of each node/economy, leading to both structural and dynamical heterogeneity among economies.
Our investigations revealed that in homogeneous (identical systems) and fully connected networks, the mean-field approximation provides an accurate description of the global dynamics, predicting a smooth transition from incoherence to complete synchronization. However, as structural or dynamical heterogeneity increases, the homogeneous mean-field framework loses validity, whereas the heterogeneous mean-field approximation remains qualitatively valid. In this heterogeneous configuration, where the systems exhibit different values of the potential GDP $d$ even within a fully connected topology, the network displays intermittent synchronization, characterized by alternating coherent and incoherent phases, with laminar phase durations following a power-law distribution with exponent $3/2$, which is in agreement with universal scaling laws for chaotic intermittency~\cite{platt1993off}. From an economic perspective, the observed intermittency implies that global business cycle coordination is inherently fragile and metastable. During coherent phases, the dynamic of economies evolve in unison, corresponding to periods of synchronized expansion or contraction—global booms or recessions—driven by strong interdependencies in trade, investment, and financial flows~\cite{imbs2004trade}. However, these coordinated periods are repeatedly disrupted by endogenous instabilities, even in the absence of external shocks~\cite{battiston2012liaisons}. This alternation between synchronization and desynchronization mirrors empirical evidence of intermittent convergence and divergence observed in real-world macroeconomic data.

\subsection{Limitations of the model}
Although the framework captures key mechanisms
of macroeconomic synchronization, several idealizations constrain limit its empirical scope: i) Coupling is diffusive, whereas real trade and capital flows involve delays, nonlinear saturation, and directional asymmetries (e.g., dollar dominance), which can induce phase lags and asymmetric shock propagation. ii) Potential GDP is treated as static. In reality, potential GDP $d$ evolves due to investment, technological change, and policy, potentially inducing adaptive rewiring and hysteresis in synchronization. iii) The model is deterministic. Stochastic exogenous shocks---e.g., policy shifts, geopolitical events, commodity price swings---are major drivers of desynchronization and are not accounted for. iv) The fitness-based connection rule, while analytically tractable, ignores institutional, geographic, and historical determinants of link formation (e.g., colonial ties, trade blocs), leading to topologies that may deviate from empirical networks. v) The Bouali--Camargo system, though grounded in macro feedbacks, omits labor markets, debt accumulation, and monetary policy---critical channels for business cycle transmission and crisis amplification.

In future work, we aim to extend the present framework by incorporating time-delayed and directed networks, endogenous fitness evolution, stochastic forcing, and empirically calibrated network structures. Such extensions could help assess more precisely how the fragility of global economic coordination emerges from the interplay between external disturbances and heterogeneous, chaotic interaction structures.


\section*{Acknowlegments}
TN acknowledges support from the ``Reconstruction, Resilience and Recovery of Socio-Economic Networks'' RECON-NET - EP\_FAIR\_005 - PE0000013 ``FAIR'' -
PNRR M4C2 Investment 1.3, financed by the European Union
– NextGenerationEU.\\

\appendix 
\section{The mean field model of dynamically heterogeneous network with high variability} \label{sec::app1}
\begin{figure}[htp!]
\centering
\begin{tabular}{c}
\hspace{-0.750 cm}
\includegraphics[width=0.55\textwidth]{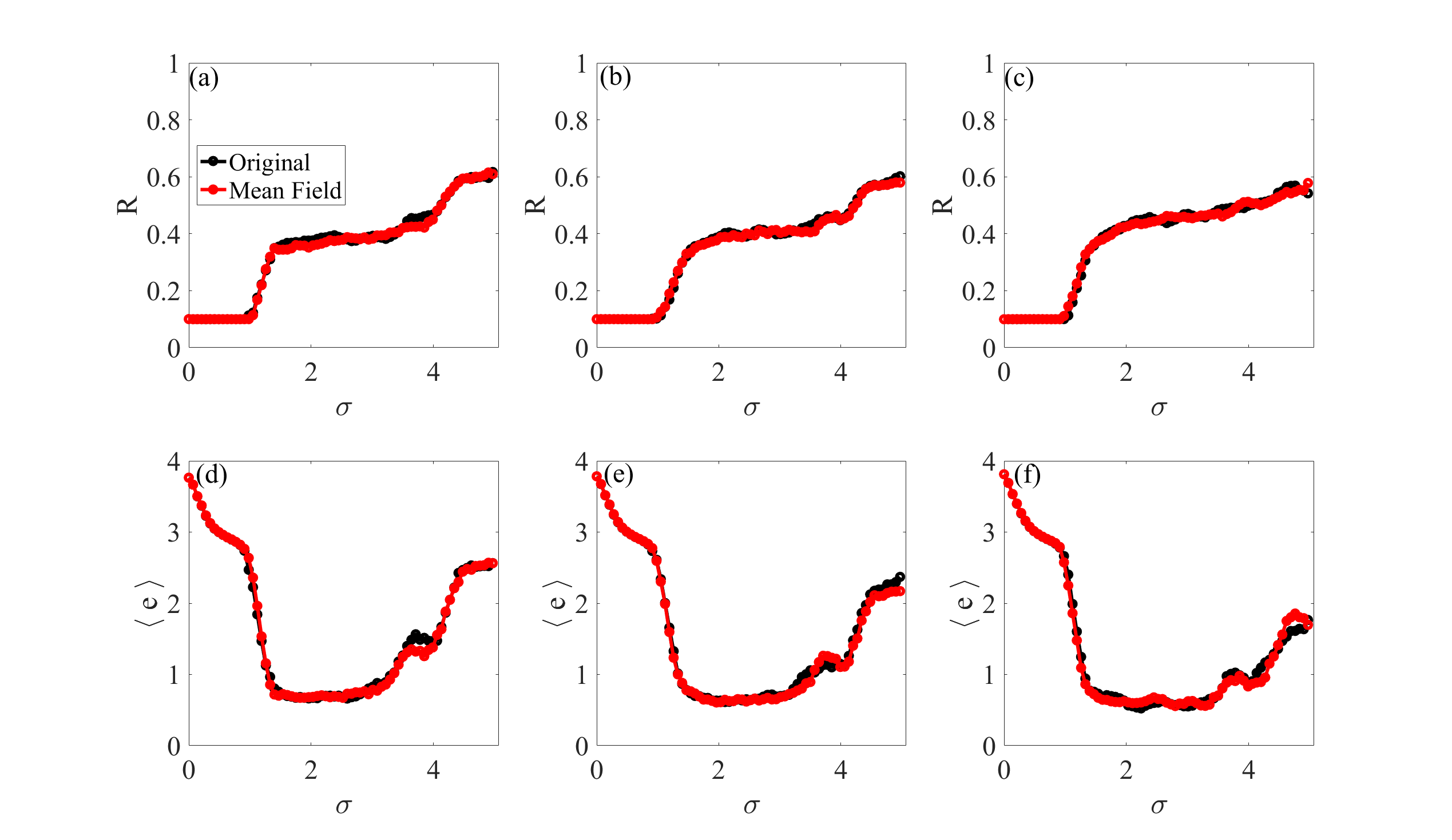}
\end{tabular}
\caption{Synchronization transitions in a heterogeneous network with a fully connected topology for $\delta = 10^{5}$, composed of nonidentical oscillators with high variability in $d$ given by $\chi=0.45$, as a function of the coupling strength~$\sigma$. Top row (a--c): average order parameter. Bottom row (d--f): average synchronization error. Columns correspond to $\bar{d}=0.8$, $1.0$, and $1.2$, respectively.}
\label{fig::Hetero_graph02_app}
\end{figure}

Fig.~\ref{fig::Hetero_graph02_app} illustrates the synchronization transition in a dynamically heterogeneous network with high variability ($\chi = 0.45$) but a homogeneous (fully connected) topology. Despite the strong dispersion in the local dynamics of individual nodes—or economies—the mean-field approximation remains valid, accurately capturing the overall collective behavior of the system. This result confirms that, in a structurally homogeneous network, the uniform coupling dominates over intrinsic dynamical diversity, effectively averaging out local fluctuations. 

\section{Limitations of the mean-field model approximation in heterogeneous networks} \label{sec::app2}
\begin{figure}[htp!]
\centering
\begin{tabular}{c}
\hspace{-0.650 cm}
\includegraphics[width=0.55\textwidth]{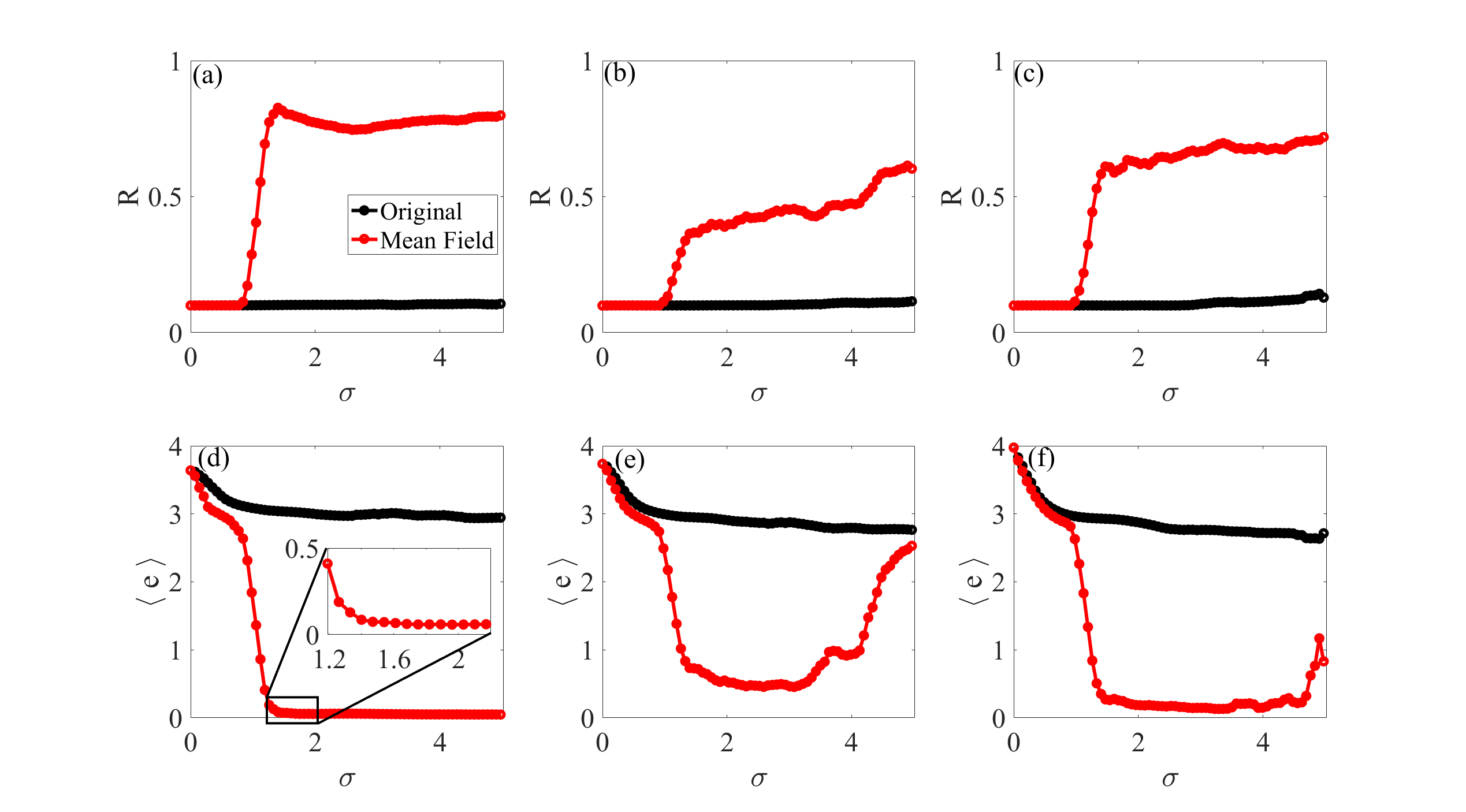}
\end{tabular}
\caption{Synchronization transitions in a heterogeneous network for $\delta = 35\times10^{-3}$, composed of nonidentical oscillators, as a function of the coupling strength~$\sigma$. Top row (a--c): average order parameter $R$. Bottom row (d--f): average synchronization error $\langle e \rangle$. The values of $\bar{d}$ are the same as those used in Fig.~\ref{fig::Hetero_graph02_app}, with variability $\chi=0.1$.
}
\label{fig::Hetero_graph01_app}
\end{figure}

The results presented in Fig.~\ref{fig::Hetero_graph01_app} highlight that increasing dynamical heterogeneity suppresses global synchronization and promotes the persistence of incoherent dynamics. As the variability in the potential GDP among nodes grows, the differences in local chaotic behavior prevent the emergence of a coherent collective state, leading instead to fragmented or partially synchronized clusters and then, the mean-field approach fails to capture the collective dynamic of the network.

\bibliography{Ref}

@article{watts1998collective,
  title={Collective dynamics of ‘small-world’networks},
  author={Watts, Duncan J and Strogatz, Steven H},
  journal={nature},
  volume={393},
  number={6684},
  pages={440--442},
  year={1998},
  publisher={Nature Publishing Group}
}

@article{barabasi2003scale,
  title={Scale-free networks},
  author={Barab{\'a}si, Albert-L{\'a}szl{\'o} and Bonabeau, Eric},
  journal={Scientific American},
  volume={288},
  number={5},
  pages={60--69},
  year={2003},
  publisher={JSTOR}
}

@article{erdds1959random,
  title={On random graphs I},
  author={ERDdS, P and R\&wi, A},
  journal={Publ. math. debrecen},
  volume={6},
  number={290-297},
  pages={18},
  year={1959}
}

@article{rives2003modular,
  title={Modular organization of cellular networks},
  author={Rives, Alexander W and Galitski, Timothy},
  journal={Proceedings of the national Academy of sciences},
  volume={100},
  number={3},
  pages={1128--1133},
  year={2003},
  publisher={The National Academy of Sciences}
}

@article{arola2022emergence,
  title={Emergence of explosive synchronization bombs in networks of oscillators},
  author={Arola-Fern{\'a}ndez, Llu{\'\i}s and Faci-L{\'a}zaro, Sergio and Skardal, Per Sebastian and Boghiu, Emanuel-Cristian and G{\'o}mez-Garde{\~n}es, Jes{\'u}s and Arenas, Alex},
  journal={Communications Physics},
  volume={5},
  number={1},
  pages={264},
  year={2022},
  publisher={Nature Publishing Group UK London}
}

@article{reina2024speed,
  title={Speed-accuracy trade-offs in best-of-n collective decision making through heterogeneous mean-field modeling},
  author={Reina, Andreagiovanni and Njougouo, Thierry and Tuci, Elio and Carletti, Timoteo},
  journal={Physical Review E},
  volume={109},
  number={5},
  pages={054307},
  year={2024},
  publisher={APS}
}

@article{bera2017chimera,
  title={Chimera states: effects of different coupling topologies},
  author={Bera, Bidesh K and Majhi, Soumen and Ghosh, Dibakar and Perc, Matja{\v{z}}},
  journal={Europhysics Letters},
  volume={118},
  number={1},
  pages={10001},
  year={2017},
  publisher={IOP Publishing}
}

@article{perez2022network,
  title={Network topological determinants of pathogen spread},
  author={P{\'e}rez-Ortiz, Mar{\'\i}a and Manescu, Petru and Caccioli, Fabio and Fern{\'a}ndez-Reyes, Delmiro and Nachev, Parashkev and Shawe-Taylor, John},
  journal={Scientific Reports},
  volume={12},
  number={1},
  pages={7692},
  year={2022},
  publisher={Nature Publishing Group UK London}
}

@article{schweitzer2009economic,
  title={Economic networks: The new challenges},
  author={Schweitzer, Frank and Fagiolo, Giorgio and Sornette, Didier and Vega-Redondo, Fernando and Vespignani, Alessandro and White, Douglas R},
  journal={science},
  volume={325},
  number={5939},
  pages={422--425},
  year={2009},
  publisher={American Association for the Advancement of Science}
}

@article{battiston2016complexity,
  title={Complexity theory and financial regulation},
  author={Battiston, Stefano and Farmer, J Doyne and Flache, Andreas and Garlaschelli, Diego and Haldane, Andrew G and Heesterbeek, Hans and Hommes, Cars and Jaeger, Carlo and May, Robert and Scheffer, Marten},
  journal={Science},
  volume={351},
  number={6275},
  pages={818--819},
  year={2016},
  publisher={American Association for the Advancement of Science}
}

@article{haldane2011systemic,
  title={Systemic risk in banking ecosystems},
  author={Haldane, Andrew G and May, Robert M},
  journal={Nature},
  volume={469},
  number={7330},
  pages={351--355},
  year={2011},
  publisher={Nature Publishing Group UK London}
}

@article{acemoglu2012network,
  title={The network origins of aggregate fluctuations},
  author={Acemoglu, Daron and Carvalho, Vasco M and Ozdaglar, Asuman and Tahbaz-Salehi, Alireza},
  journal={Econometrica},
  volume={80},
  number={5},
  pages={1977--2016},
  year={2012},
  publisher={Wiley Online Library}
}

@article{fagiolo2010evolution,
  title={The evolution of the world trade web: a weighted-network analysis},
  author={Fagiolo, Giorgio and Reyes, Javier and Schiavo, Stefano},
  journal={Journal of Evolutionary Economics},
  volume={20},
  number={4},
  pages={479--514},
  year={2010},
  publisher={Springer}
}

@article{bardoscia2021physics,
  title={The physics of financial networks},
  author={Bardoscia, Marco and Barucca, Paolo and Battiston, Stefano and Caccioli, Fabio and Cimini, Giulio and Garlaschelli, Diego and Saracco, Fabio and Squartini, Tiziano and Caldarelli, Guido},
  journal={Nature Reviews Physics},
  volume={3},
  number={7},
  pages={490--507},
  year={2021},
  publisher={Nature Publishing Group UK London}
}

@article{colon2017economic,
  title={Economic networks: Heterogeneity-induced vulnerability and loss of synchronization},
  author={Colon, C{\'e}lian and Ghil, Michael},
  journal={Chaos: An Interdisciplinary Journal of Nonlinear Science},
  volume={27},
  number={12},
  year={2017},
  publisher={AIP Publishing}
}

@article{bouali1999feedback,
  title={Feedback loop in extended Van der Pol's equation applied to an economic model of cycles},
  author={Bouali, Safieddine},
  journal={International Journal of Bifurcation and Chaos},
  volume={9},
  number={04},
  pages={745--756},
  year={1999},
  publisher={World Scientific}
}

@article{camargo2022synchronization,
  title={Synchronization and bifurcation in an economic model},
  author={Camargo, Victor E and Amaral, Amaury S and Crepaldi, Ant{\^o}nio F and Ferreira, Fernando F},
  journal={Chaos: An Interdisciplinary Journal of Nonlinear Science},
  volume={32},
  number={10},
  year={2022},
  publisher={AIP Publishing}
}

@article{garlaschelli2004fitness,
  title={Fitness-dependent topological properties of the world trade web},
  author={Garlaschelli, Diego and Loffredo, Maria I},
  journal={Physical Review Letters},
  volume={93},
  number={18},
  pages={188701},
  year={2004},
  publisher={APS}
}

@article{fiedler1973algebraic,
  title={Algebraic connectivity of graphs},
  author={Fiedler, Miroslav},
  journal={Czechoslovak mathematical journal},
  volume={23},
  number={2},
  pages={298--305},
  year={1973},
  publisher={Institute of Mathematics, Academy of Sciences of the Czech Republic}
}

@book{jackson2003strategic,
  title={A strategic model of social and economic networks},
  author={Jackson, Matthew O and Wolinsky, Asher},
  year={2003},
  publisher={Springer}
}

@article{kuramoto2002coexistence,
  title={Coexistence of coherence and incoherence in nonlocally coupled phase oscillators},
  author={Kuramoto, Yoshiki and Battogtokh, Dorjsuren},
  journal={arXiv preprint cond-mat/0210694},
  year={2002}
}

@article{boccaletti2002synchronization,
  title={The synchronization of chaotic systems},
  author={Boccaletti, Stefano and Kurths, J{\"u}rgen and Osipov, Grigory and Valladares, Daniel L and Zhou, Changsong},
  journal={Physics Reports},
  volume={366},
  number={1--2},
  pages={1--101},
  year={2002},
  publisher={Elsevier}
}

@article{platt1993off,
  title={On-off intermittency: A mechanism for bursting},
  author={Platt, NSEA and Spiegel, EA and Tresser, C},
  journal={Physical Review Letters},
  volume={70},
  number={3},
  pages={279},
  year={1993},
  publisher={APS}
}

@article{koronovskii2024intermittent,
  title={Intermittent generalized synchronization and modified system approach: Discrete maps},
  author={Koronovskii, Alexey A and Moskalenko, Olga I and Selskii, Anton O},
  journal={Physical Review E},
  volume={109},
  number={6},
  pages={064217},
  year={2024},
  publisher={APS}
}

@article{wolf1985determining,
  title={Determining Lyapunov exponents from a time series},
  author={Wolf, Alan and Swift, Jack B and Swinney, Harry L and Vastano, John A},
  journal={Physica D: nonlinear phenomena},
  volume={16},
  number={3},
  pages={285--317},
  year={1985},
  publisher={Elsevier}
}

@article{mirollo1990amplitude,
  title={Amplitude death in an array of limit-cycle oscillators},
  author={Mirollo, Renato E and Strogatz, Steven H},
  journal={Journal of Statistical Physics},
  volume={60},
  number={1},
  pages={245--262},
  year={1990},
  publisher={Springer}
}

@article{montbrio2020exact,
  title={Exact mean-field theory explains the dual role of electrical synapses in collective synchronization},
  author={Montbri{\'o}, Ernest and Paz{\'o}, Diego},
  journal={Physical Review Letters},
  volume={125},
  number={24},
  pages={248101},
  year={2020},
  publisher={APS}
}

@article{rosenblum1996phase,
  title={Phase synchronization of chaotic oscillators},
  author={Rosenblum, Michael G and Pikovsky, Arkady S and Kurths, J{\"u}rgen},
  journal={Physical Review Letters},
  volume={76},
  number={11},
  pages={1804},
  year={1996},
  publisher={APS}
}

@article{pikovsky2001synchronization,
  title={Synchronization},
  author={Pikovsky, Arkady and Rosenblum, Michael and Kurths, J{\"u}rgen},
  journal={Cambridge university press},
  volume={12},
  year={2001},
  publisher={Cambridge university press}
}

@techreport{ikeda2013synchronization,
  title={Synchronization and the coupled oscillator model in international business cycles},
  author={Ikeda, Yuichi and Aoyama, Hideaki and Yoshikawa, Hiroshi},
  year={2013},
  institution={RIETI Discussion Paper October 13-E}
}

@article{boguna2003absence,
  title={Absence of epidemic threshold in scale-free networks with degree correlations},
  author={Bogun{\'a}, Mari{\'a}n and Pastor-Satorras, Romualdo and Vespignani, Alessandro},
  journal={Physical review letters},
  volume={90},
  number={2},
  pages={028701},
  year={2003},
  publisher={APS}
}

@article{caldarelli2002scale,
  title={Scale-free networks from varying vertex intrinsic fitness},
  author={Caldarelli, Guido and Capocci, Andrea and De Los Rios, Paolo and Munoz, Miguel A},
  journal={Physical review letters},
  volume={89},
  number={25},
  pages={258702},
  year={2002},
  publisher={APS}
}

@article{imbs2004trade,
  title={Trade, finance, specialization, and synchronization},
  author={Imbs, Jean},
  journal={Review of economics and statistics},
  volume={86},
  number={3},
  pages={723--734},
  year={2004},
  publisher={MIT Press 238 Main St., Suite 500, Cambridge, MA 02142-1046, USA journals~…}
}

@article{battiston2012liaisons,
  title={Liaisons dangereuses: Increasing connectivity, risk sharing, and systemic risk},
  author={Battiston, Stefano and Gatti, Domenico Delli and Gallegati, Mauro and Greenwald, Bruce and Stiglitz, Joseph E},
  journal={Journal of economic dynamics and control},
  volume={36},
  number={8},
  pages={1121--1141},
  year={2012},
  publisher={Elsevier}
}

@article{molloy1995critical,
  title={A critical point for random graphs with a given degree sequence},
  author={Molloy, Michael and Reed, Bruce},
  journal={Random structures \& algorithms},
  volume={6},
  number={2-3},
  pages={161--180},
  year={1995},
  publisher={Wiley Online Library}
}

@article{caprioglio2024emergence,
  title={Emergence of metastability in frustrated oscillatory networks: the key role of hierarchical modularity},
  author={Caprioglio, Enrico and Berthouze, Luc},
  journal={Frontiers in Network Physiology},
  volume={4},
  pages={1436046},
  year={2024},
  publisher={Frontiers Media SA}
}

@article{simo2021chimera,
  title={Chimera states in a neuronal network under the action of an electric field},
  author={Simo, Ga{\"e}l R and Njougouo, Thierry and Aristides, RP and Louodop, Patrick and Tchitnga, Robert and Cerdeira, Hilda A},
  journal={Physical Review E},
  volume={103},
  number={6},
  pages={062304},
  year={2021},
  publisher={APS}
}

@article{muolo2024phase,
  title={Phase chimera states on nonlocal hyperrings},
  author={Muolo, Riccardo and Njougouo, Thierry and Gambuzza, Lucia Valentina and Carletti, Timoteo and Frasca, Mattia},
  journal={Physical Review E},
  volume={109},
  number={2},
  pages={L022201},
  year={2024},
  publisher={APS}
}

@inproceedings{njougouo2023heterogeneous,
  title={Heterogeneous mean-field analysis of best-of-n decision making in networks with zealots},
  author={Njougouo, Thierry and Carletti, Timoteo and Reina, Andreagiovanni and Tuci, Elio},
  booktitle={Italian Workshop on Artificial Life and Evolutionary\\ Computation},
  pages={339--351},
  year={2023},
  organization={Springer}
}

@article{meylahn2024opinion,
  title={Opinion dynamics beyond social influence},
  author={Meylahn, Benedikt V and Searle, Christa},
  journal={Network Science},
  volume={12},
  number={4},
  pages={339--365},
  year={2024},
  publisher={Cambridge University Press}
}

@article{cencetti2023distinguishing,
  title={Distinguishing simple and complex contagion processes on networks},
  author={Cencetti, Giulia and Contreras, Diego Andr{\'e}s and Mancastroppa, Marco and Barrat, Alain},
  journal={Physical Review Letters},
  volume={130},
  number={24},
  pages={247401},
  year={2023},
  publisher={APS}
}

@article{saberi2020simple,
  title={A simple contagion process describes spreading of traffic jams in urban networks},
  author={Saberi, Meead and Hamedmoghadam, Homayoun and Ashfaq, Mudabber and Hosseini, Seyed Amir and Gu, Ziyuan and Shafiei, Sajjad and Nair, Divya J and Dixit, Vinayak and Gardner, Lauren and Waller, S Travis and others},
  journal={Nature communications},
  volume={11},
  number={1},
  pages={1616},
  year={2020},
  publisher={Nature Publishing Group UK London}
}

@article{njougouo2020effects,
  title={Effects of intermittent coupling on synchronization},
  author={Njougouo, Thierry and Simo, Ga{\"e}l R and Louodop, Patrick and Fotsin, Hilaire and Talla, Pierre K},
  journal={Chaos, Solitons \& Fractals},
  volume={139},
  pages={110082},
  year={2020},
  publisher={Elsevier}
}

@article{njougouo2020dynamics,
  title={Dynamics of multilayer networks with amplification},
  author={Njougouo, Thierry and Camargo, Victor and Louodop, Patrick and Fagundes Ferreira, Fernando and Talla, Pierre K and Cerdeira, Hilda A},
  journal={Chaos: An Interdisciplinary Journal of Nonlinear Science},
  volume={30},
  number={12},
  year={2020},
  publisher={AIP Publishing}
}

@article{zheng2000generalized,
  title={Generalized synchronization versus phase synchronization},
  author={Zheng, Zhigang and Hu, Gang},
  journal={Physical Review E},
  volume={62},
  number={6},
  pages={7882},
  year={2000},
  publisher={APS}
}
\end{document}